\pdfoutput=1
\documentclass[sn-mathphys]{sn-jnl}

\usepackage{bbm,graphicx,color,euscript,comment,mathtools,natbib}
\graphicspath{{paperfigs/}}
\usepackage{cleveref}
\crefname{figure}{Fig.}{Figs.}
\usepackage{amsmath,amssymb}
\usepackage{caption}
\usepackage{subcaption}
\usepackage{bm}
\usepackage[bb=boondox, bbscaled=1.06,cal=euler,scr=boondoxo,scrscaled=1.04,frak=euler,frakscaled=1.05]{mathalpha}

\def\rv#1{\textcolor{black}{#1}}

\newcommand{\sfN}{\mathcal{N}}

\newcommand{\bfb}{\mathbf{b}}
\newcommand{\bfn}{\mathbf{n}}
\newcommand{\bft}{\mathbf{t}}
\newcommand{\bfm}{\mathbf{m}}

\newcommand{\bfx}{\mathbf{x}}
\newcommand{\bfy}{\mathbf{y}}
\newcommand{\bfz}{\mathbf{z}}

\DeclareMathOperator {\diag}{diag}

\def\var#1{\mathbb{V}\text{ar}\left[#1\right]}
\def\githublink{\url{https://github.com/michael-c-brennan/DFXMTools}}

\def\forward{\mathcal{F}}

\def\xlab{x_{\ell}}
\def\ylab{y_{\ell}}
\def\zlab{z_{\ell}}

\def\xdis{x_{d}}
\def\ydis{y_{d}}
\def\zdis{z_{d}}

\def\xdisvec{\bfx_{d}}
\def\ydisvec{\bfy_{d}}
\def\zdisvec{\bfz_{d}}

\def\xsample{x_{s}}
\def\ysample{y_{s}}
\def\zsample{z_{s}}

\def\pos{\xi}
\def\mappos{\pos^{\text{MAP}}}
\def\truepos{\pos^{\text{true}}}

\def\obs{\mathbf{I}}
\def\noisefreeobs{\obs^{\text{nf}} }
\def\trueobs{\obs^{\text{obs}}}
\def\pixelvalue{I_{n}}

\def\numpixels{N_{\text{pixels}}}
\def\NA{N\!A}

\def\argmax{\operatornamewithlimits{argmax}}
\def\argmin{\operatornamewithlimits{argmin}}

\jyear{2021}%

\theoremstyle{thmstyleone}
\theoremstyle{thmstyletwo}%

\theoremstyle{thmstylethree}%

\begin{document}

\title[Analytical Methods for Superresolution Dislocation Identification in DFXM]{Analytical Methods for Superresolution Dislocation Identification in Dark-Field X-ray Microscopy}


\author*[1]{\fnm{Michael C.} \sur{Brennan}}\email{mcbrenn@mit.edu}

\author[2]{\fnm{Marylesa} \sur{Howard}}\email{howardmm@nv.doe.gov}

\author[1]{\fnm{Youssef} \sur{Marzouk}}\email{ymarz@mit.edu}

\author*[3,4,5]{\fnm{Leora E.} \sur{Dresselhaus-Marais}}\email{leoradm@stanford.edu}

\affil[1]{\orgdiv{Department of Aeronautics and Astronautics, Center for Computational Science and Engineering}, \orgname{Massachusetts Institute of Technology}, \orgaddress{\street{77 Massachusetts Ave}, \city{Cambridge}, \postcode{02139}, \state{MA}, \country{USA}}}

\affil[2]{\orgdiv{Experimental Operations}, \orgname{Nevada National Security Site}, \orgaddress{\street{232 Energy Way}, \city{North Las Vegas}, \postcode{89030}, \state{NV}, \country{USA}}}

\affil[3]{\orgdiv{Department of Materials Science and Engineering}, \orgname{Stanford University}, \orgaddress{\street{476 Lomita Mall}, \city{Stanford}, \postcode{94305}, \state{CA}, \country{USA}}}

\affil[4]{\orgdiv{Department of Photon Science}, \orgname{SLAC National Accelerator Laboratory}, \orgaddress{\street{2575 Sand Hill Rd}, \city{Menlo Park}, \postcode{94025}, \state{CA}, \country{USA}}}

\affil[5]{\orgdiv{Physics Division}, \orgname{Lawrence Livermore National Laboratory}, \orgaddress{\street{7000 East Ave}, \city{Livermore}, \postcode{94550}, \state{CA}, \country{USA}}}

\abstract{We develop several inference methods to estimate the position of dislocations from images generated using dark-field X-ray microscopy (DFXM)---achieving superresolution accuracy and principled uncertainty quantification. Using the framework of Bayesian inference, we incorporate models of the DFXM contrast mechanism and detector measurement noise, along with initial position estimates, into a statistical model coupling DFXM images with the dislocation position of interest. We motivate several position estimation and uncertainty quantification algorithms based on this model. We then demonstrate the accuracy of our primary estimation algorithm on synthetic realistic DFXM images of edge dislocations in single crystal aluminum. We conclude with a discussion of our methods' impact on future dislocation studies and possible future research avenues.}


\maketitle


\section{Introduction}
\label{sec:intro}
Dislocations are ubiquitous in studies of the mechanical properties of crystalline materials \cite{hull2011introduction}. For roughly 70 years, dislocations have been studied with electron microscopes that resolve the relevant dynamics and interactions across angstrom to nanometer length scales. As electron microscopes can only penetrate up to $2 \ \mu$m, they can only characterize surfaces or subsurface dynamics in very thin films, which have been shown to exhibit surface and size effects non-typical of their bulk counterparts \cite{Kubin2011}. As such, our ability to test models for crystal plasticity are limited for macroscopic materials. X-ray imaging methods like X-ray topography and radiography can directly measure the relevant subsurface dynamics, but have long suffered from ambiguity in interpretation for high dislocation-density metals, due to their limitations in spatial resolution \cite{hull2011introduction, Danilewsky2020}. 

Dark field X-ray microscopy (DFXM) was developed over the past decade as a new full-field imaging method that images deformations that can be hundreds of micrometers beneath any surface, with up to $30$--$150$ nm resolution \cite{kutsal2019esrf}. Using an X-ray objective lens along the diffracted beam, DFXM directly maps the orientation and strain (i.e., a projection of the displacement gradient tensor field) of crystalline materials \cite{simons2015dark,poulsen2017x,poulsen2020multi}. Compared to other imaging methods, DFXM has several attractive characteristics. DFXM uses high energy X-rays, which can penetrate more deeply than a micron, and hence allows for imaging within bulk materials. High energy X-rays also have large extinction lengths, which limits dynamical diffraction (i.e., multiple scattering) and simplifies experiment design and interpretation. DFXM has been employed in the study of ferroelectrics \cite{simons2018long} and biominerals \cite{cook2018insights}, and recently has been applied to analyze populations of dislocations \cite{jakobsen2019mapping,poulsen2020geometrical,gonzalez2020methods,dresselhaus2021situ}. 

While DFXM holds significant promise to characterize \textit{bulk} dislocation-mediated plasticity, its application has still been limited to studies that map either the \textit{density} of dislocations or the dynamics of dislocations spread by more than $1 \, \mu$m. As DFXM's spatial resolution is significantly higher (i.e., finer) than the distance between dislocations, this raises the question: \emph{What has prevented DFXM from accessing the $\approx$100 nm separated boundaries that are routinely observed in TEM?} The challenge thus far has been the high sensitivity of DFXM's measurements. Since DFXM maps the displacement gradient tensor field $\nabla \bm{u}(x,y,z)$ with resolution on the order $10^{-5}$, dislocations appear on the order of a micron in size in images, which is far larger than it would be via TEM and significantly larger than the angstrom dimension of the dislocation core \cite{poulsen2020geometrical}. 

In this work, we extend conventional DFXM analysis to TEM-scale accuracy using statistical inference methods. We develop a numerical method that uses Bayesian inference to improve the accuracy of our prediction of the dislocation position based on the physics of DFXM contrast and CMOS noise processes. We apply our new method to simulated DFXM images of edge dislocations in single-crystal aluminum, as this well-defined system allows us to define a ground truth for the dislocation’s position, against which we can evaluate the accuracy of our method. To ensure that our results are robust to the more complex features in experimental data, we add a measurement noise model to the forward model recently developed in \cite{poulsen2020geometrical}---a formalism that may easily be extended to other materials. Our approach achieves an accuracy that exceeds \textit{even the best $30$ nm resolution of the imaging optics.} Our inference methods also offer a complete quantification of uncertainty in the dislocation position, along the characteristic vectors defining the dislocation (i.e., the slip plane normal and Burger's vector). After demonstrating our approach, we provide guidance on use cases for each of our algorithms, and present an outlook on how to extend our methods to interpret experimental data collected at synchrotrons. We expect our inferential approach to open a new range of possibilities for characterizing subsurface dislocation dynamics---with important opportunities for structural mechanics, dislocation theory, thermal transport, and beyond.  


\section{Simulating DFXM images}\label{sec:model}
Our position estimation and uncertainty quantification algorithms require the ability to simulate the imaging process for the material being sampled, the configuration of the microscope, and parameters of the incident X-ray beam.
In this section we describe our method of DFXM image simulation, which builds upon a model described previously in \cite{poulsen2020geometrical}. This model uses continuum mechanics to describe the material and a wave-optics formalism to simulate idealized noise-free DFXM images, $\noisefreeobs$, given the experimental parameters describing the sample and experimental setup. In our use of this model, we denote the position of the dislocation of interest for these predictions as $\pos$; we denote our simulation process as $\forward$, which we call the \emph{forward model}, such that $\noisefreeobs = \forward(\pos)$. We omit the full model derivation in this work, and instead summarize its key steps and simplifying assumptions---to contextualize its scope and highlight new elements in the present implementation.

\subsection{The forward model}
\label{sec:model:forwardmodel}

The authors of \cite{poulsen2020geometrical} express the intensity of X-ray light that reaches all points on the detector as the convolution of a micromechanical model with an instrumental resolution function. The micromechanical model describes the un-deformed and deformed regions of the crystal lattice, taking into account the space-group and associated local deformations. Using a continuum mechanics approach, this model relates the displacement gradient tensor field $\nabla \bm{u}$ to reciprocal lattice vectors, $\bf{q}$, at each position in the sample at the microscale. The micromechanical model depends on the attributes of the material one is sampling and on the configuration of the microscope. In this case, we describe the dislocations using elastic theory, as described in \cite{hull2011introduction}, because this theory adequately describes the long-range distortion fields; in more complex systems, it may be extended to more detailed models, so long as they can be voxelized into $\nabla \bm{u}$.

The instrumental resolution function accounts for the experimental specifications of the instrument, including the goniometer angles to orient the sample, the parameters of the incident X-ray beam, and the specified imaging components. 
The authors of \cite{poulsen2020geometrical} assemble their resolution function as a histogram of Monte Carlo-simulated incident rays that diffract and transmit through the imaging optics to reach the specified detector coordinates. The resolution function can be computed independently of the specific dislocation type, character, and position, though the instrument parameters must be configured to ensure that they contribute nonzero intensity contrast from the dislocation of interest. 
The noise-free intensity for each pixel is defined by integrating over the real-space voxel in the sample, assuming a gauge volume set by the projection angle of the diffraction. For our numerical implementation of the forward model, see \githublink. 

For the purposes of this work, we make a small generalization to the model in \cite{poulsen2020geometrical} in order to handle the case of multiple dislocations. Let $\nabla \bm{u}_{\pos_i}$ denote the displacement gradient tensor field emanating from a dislocation core located at position $\pos_i$. In this case, the total displacement gradient tensor induced by a set of dislocations at positions $\pos = \{\pos_1, \dots, \pos_D\}$ is equal to the sum of the individual tensor fields,
$$
\nabla \bm{u} = \sum_{i = 1}^D \nabla \bm{u}_{\pos_i}.
$$

\subsection{Optical parameters and material constants}
\label{sec:model:parameters}

Above we introduce the forward model as it is considered in setting up experiments for effective implementation and contrast on desired dislocations, as described in \cite{poulsen2020geometrical}. While this was important to set up our \emph{initial} modeling framework for this study, we emphasize that our approach in this work views this model through a different perspective. Our \emph{analysis-focused} approach takes the microscope parameters simply as inputs that are set by the experiment being analyzed. As such, we assume that the free variable is the micromechanical model describing the defects that \emph{might} describe a dataset/evaluation image. In this way, our change in perspective on this model allows us to focus on the \emph{interpretation} of the results for materials science.

We now list and categorize the inputs that define the forward model.
The parameters describing the material and microscope configuration together define the micromechanical model, while the characteristics of the X-ray beam and microscope configuration define the instrumental resolution function.
The material attributes are defined by the lattice parameters of the material, $a=b=c=4.0479 \text{\AA}$ and $\alpha=\beta=\gamma=90^{\circ}$ for FCC aluminum; the length of the Burger's vector for the relevant dislocation system, $\|\bfb\|$; the Poisson ratio, $\nu$; and the $d$-spacing for the relevant $hk\ell$ vector of the diffracting plane of the crystal. 
We use the diffraction angle $2\theta$ of a given deformation state to define the bounds on all goniometer angles, i.e., the angles that we use to change the contrast mechanism/diffraction condition by re-orienting the sample ($\phi$, $\chi$, $\omega$). The resolution function also includes the parameters of the incident X-ray beam, which has a characteristic profile in the $\ylab,\zlab$ directions and its vertical and horizontal divergences ($\Delta \zeta_v$, $\Delta \zeta_h$), and Gaussian energy variation $\Delta E/E$. We assume the incident beam's intensity to be a top-hat function in the $\ylab$ direction, and Gaussian in the $\zlab$ direction with RMS $\Delta \zlab$. This assumption is consistent with the aperture and 1D focusing optics used in the synchrotron experiments, respectively. The imaging distances that determine the magnification in the imaging condition are also inputs to the resolution function: specifically the distance from the sample to the entry point of the objective, $d_1$; the physical aperture, $D$; and the numerical aperture, $\NA$. 

\begin{figure}[t]
    \centering
    \includegraphics[width=\linewidth,trim=0cm 0cm 0cm 0cm,clip]{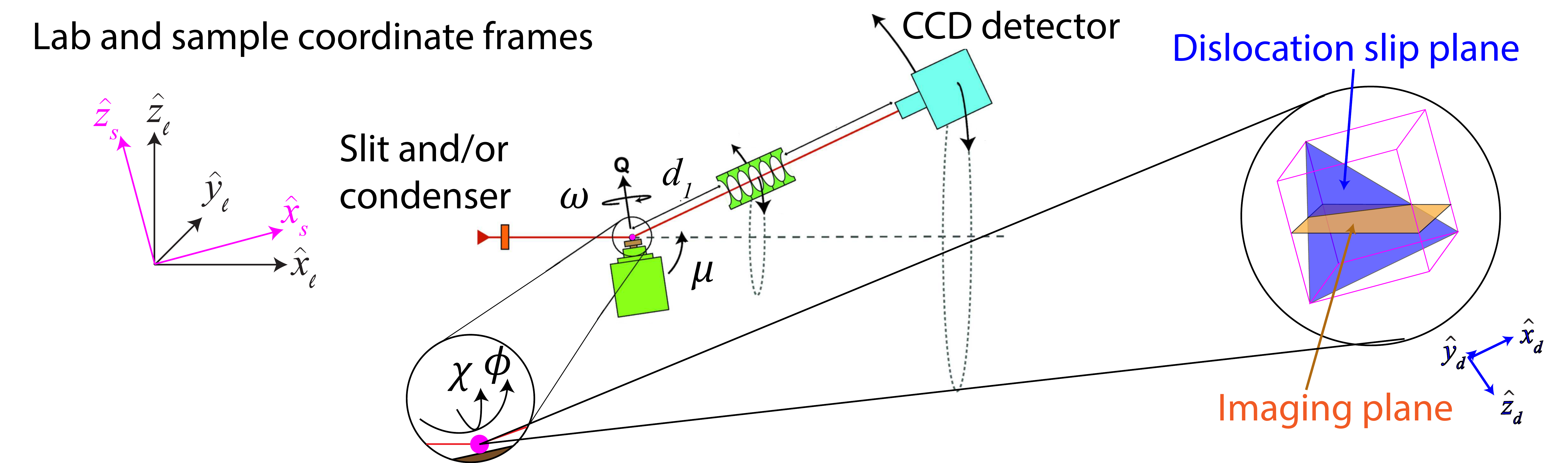}
    \caption{The geometry of the dark-field X-ray microscope at ID06 at the ESRF, which defines several coordinate frames relevant to our work. The laboratory coordinate system, $(\xlab,\ylab,\zlab)$ is aligned with the image plane set by the X-ray beam. The sample coordinate system $(\xsample,\ysample,\zsample)$, is defined as a rotation of the lab coordinates by the goniometer angles (base tilt $\mu$, rotation $\omega$, and tilts $\chi$ and $\phi$) and captures the configuration of the microscope. The dislocation coordinate system $(\xdis,\ydis,\zdis)$ is defined by the Burgers vector $\xdisvec = \bfb/\|\bfb\|$, slip plane normal vector $\ydisvec = \bfn/\|\bfn\|$, and line vector $\zdisvec = \bft/\|\bft\|$, and is shown in relation to the sample frame and image plane. This figure is adapted from \cite{poulsen2017x}.
}
    \label{fig:esrf-setup}
\end{figure}

\begin{figure}[t]
	\centering
	\begin{minipage}{.24\linewidth}
		\begin{subfigure}[b]{\linewidth}
			\includegraphics[width=\linewidth,trim=3.5cm 1cm 2cm 1cm,clip]{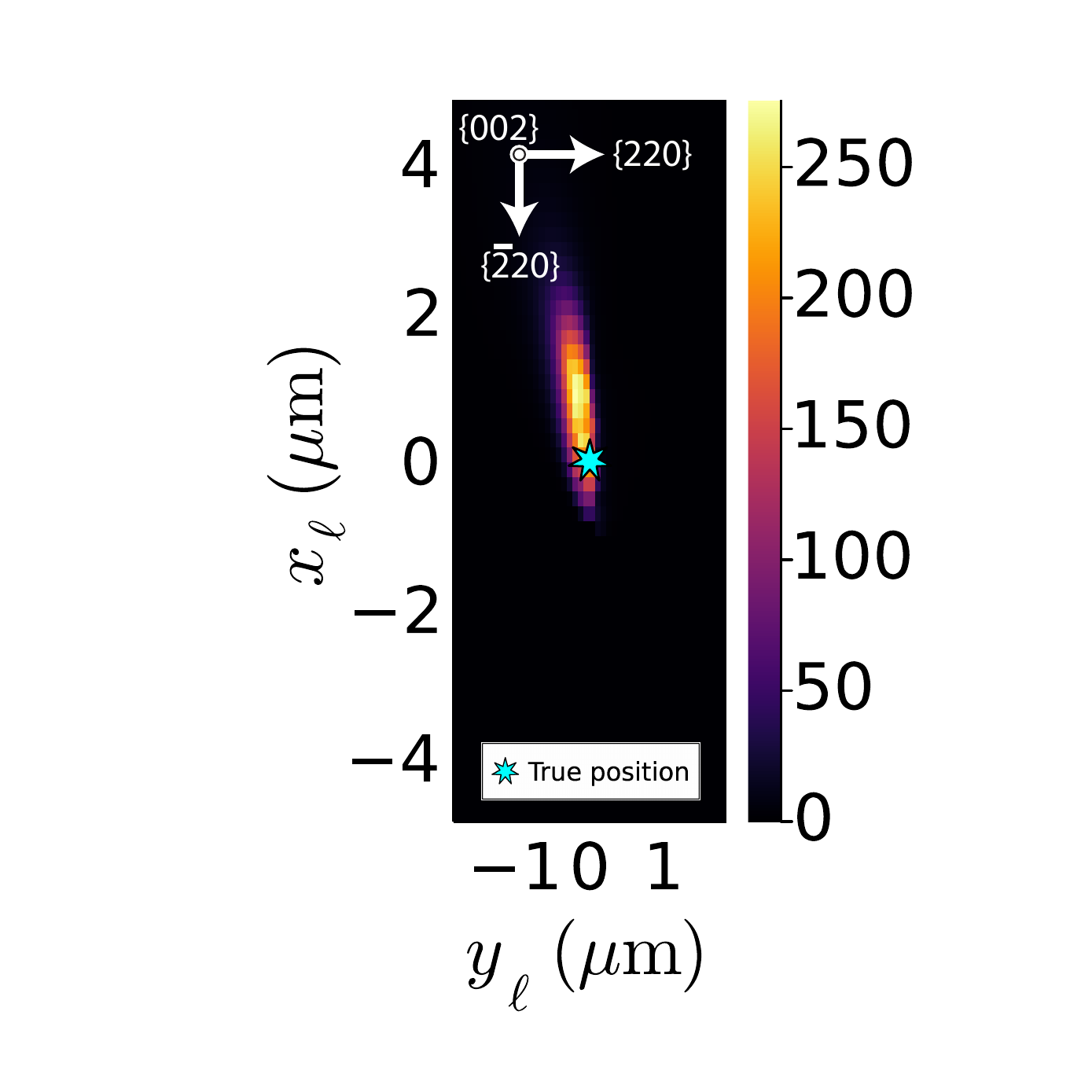}
			\caption{}
			\label{fig:1-disloc}
		\end{subfigure}
	\end{minipage}
	\begin{minipage}{.24\linewidth}
		\begin{subfigure}[b]{\linewidth}
			\includegraphics[width=\linewidth,trim=3.5cm 1cm 2cm 1cm,clip]{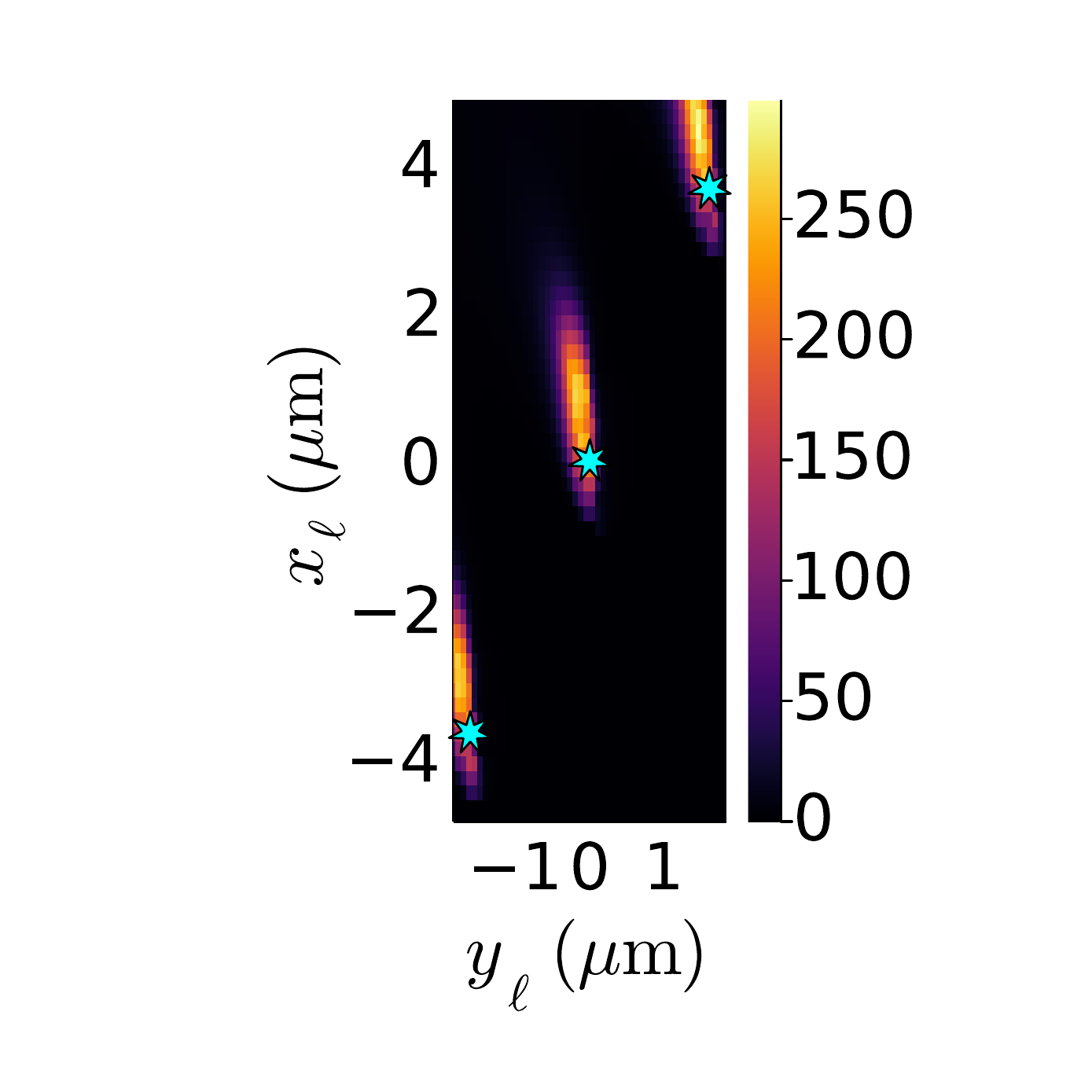}
			\caption{}
			\label{fig:3-disloc}
		\end{subfigure}
	\end{minipage}
	\begin{minipage}{.24\linewidth}
		\begin{subfigure}[b]{\textwidth}
			\includegraphics[width=\linewidth,trim=3.5cm 1cm 2cm 1cm,clip]{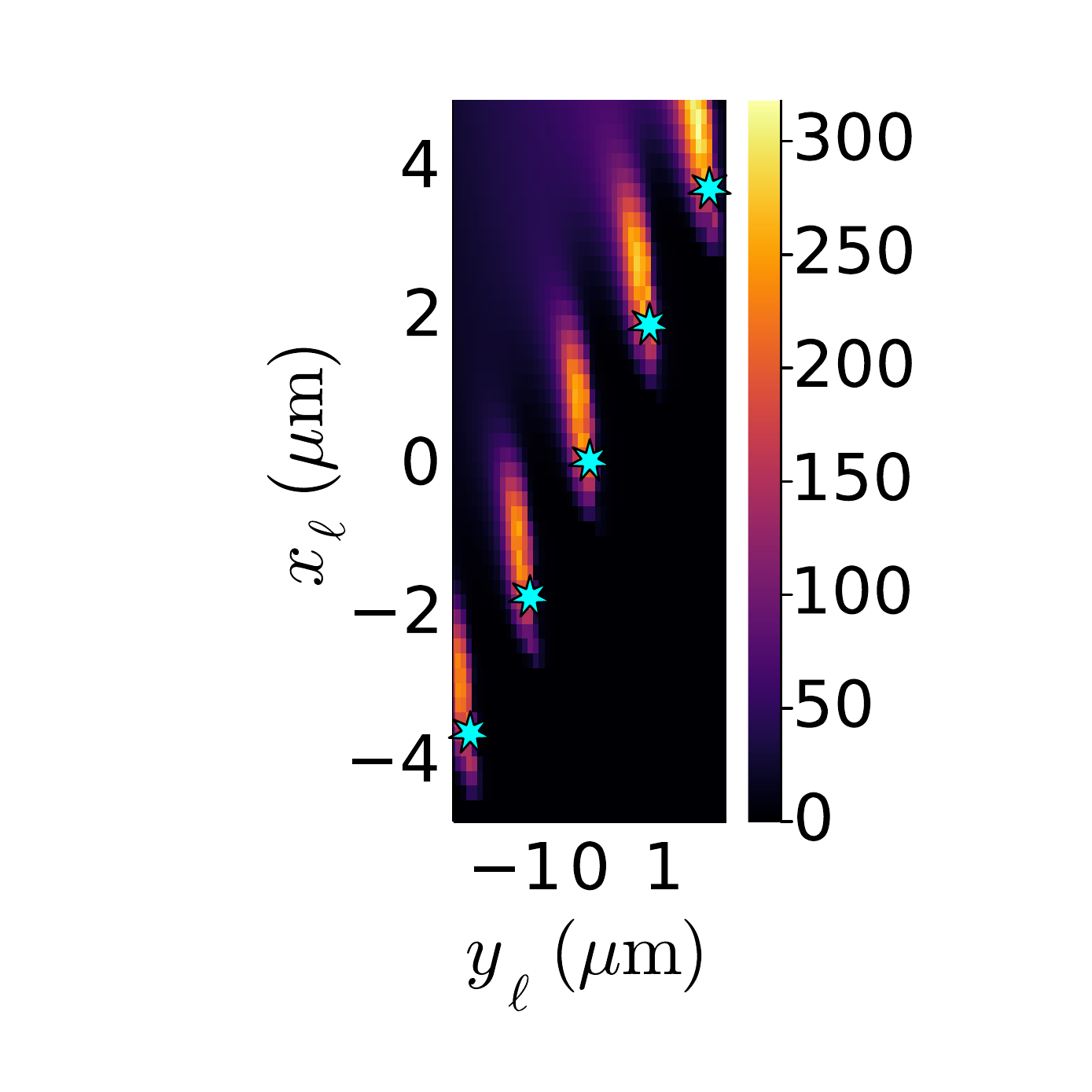}
			\caption{}
			\label{fig:5-disloc}
		\end{subfigure}	
	\end{minipage}
	\begin{minipage}{.24\linewidth}
		\begin{subfigure}[b]{\linewidth}
			\includegraphics[width=\linewidth,trim=3.5cm 1cm 2cm 1cm,clip]{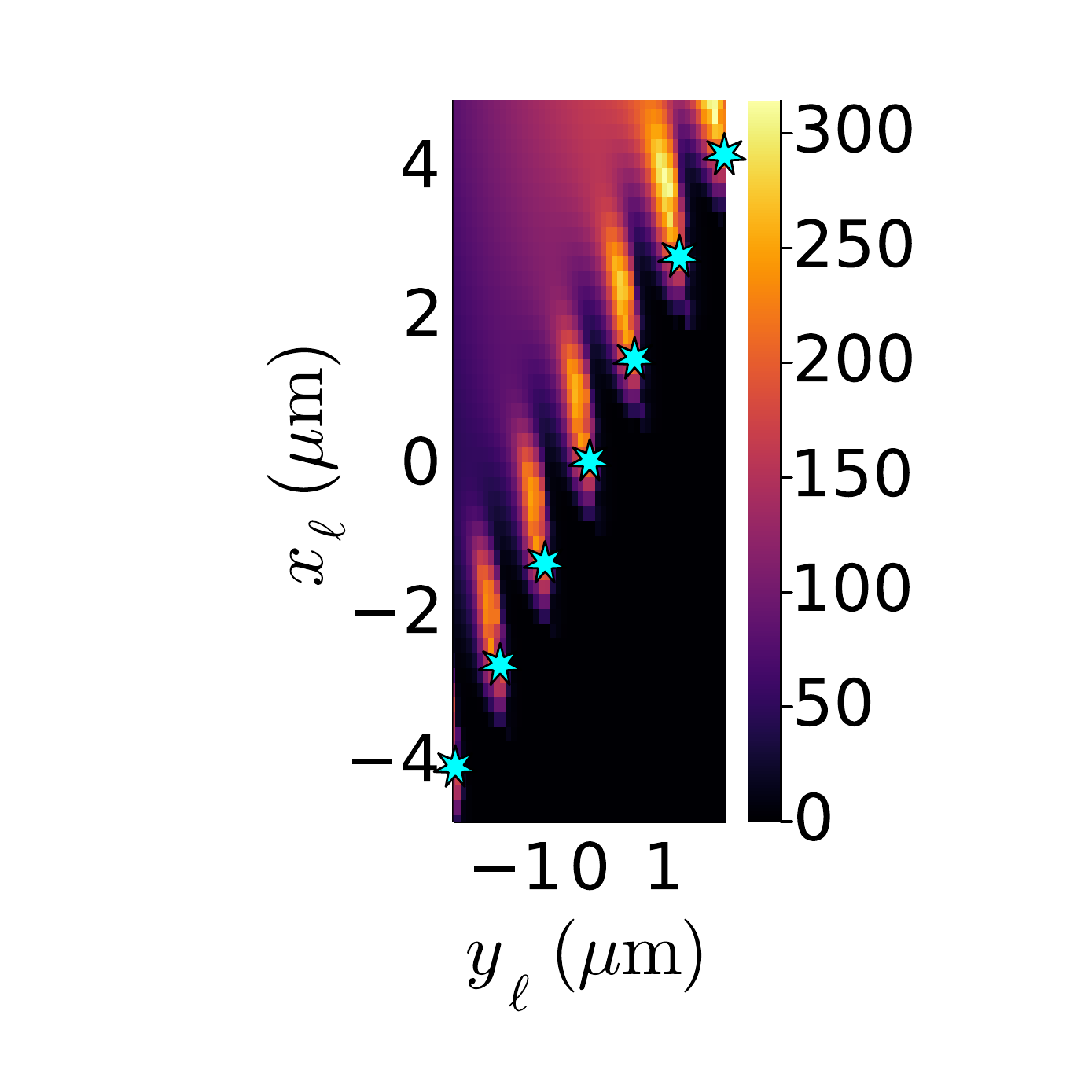}
			\caption{}
			\label{fig:7-disloc}
		\end{subfigure}	
		
	\end{minipage}
	\caption{
            Simulated images $\noisefreeobs$ of edge dislocations with Burgers vector $\bfb = [1 \ 1 \ 0]$ and slip plane normal $\bfn = [1 \ \bar{1}\ 1]$. \Cref{fig:1-disloc} shows a lone dislocation. \Cref{fig:3-disloc} shows dislocations spaced $4$ $\mu$m apart, \cref{fig:5-disloc} shows dislocations spaced $2$ $\mu$m apart, and \cref{fig:7-disloc} shows dislocations spaced $1.5$ $\mu$m apart. The dislocations are aligned along their shared Burgers vector, implying that they pack into a stabilizing tilt boundary, as discussed in \cite{dresselhaus2021situ}.
}
	\label{fig:multiple-dislocations}
\end{figure}

\Cref{fig:esrf-setup} shows the geometry of the dark field X-ray microscope at ID06 at the European Synchrotron Radiation Facility (ESRF), and provides a visual representation of the parameters discussed above when applicable. For further discussion of how each parameter enters into the physical DFXM model, see \cite{poulsen2020geometrical}. The values for all parameters used in this work replicate the experiment presented in \cite{dresselhaus2021situ}, and are included in \Cref{tab:setup}. All distribution widths (e.g., the vertical divergence $\Delta \zeta_v$) are given in terms of the RMS value. \Cref{fig:multiple-dislocations} shows simulated images for different configurations of edge dislocations, using the simulation parameters defined in \Cref{tab:setup}.

\begin{table}[h!]
	\centering
	\caption{Forward model inputs}
	\begin{tabular}{ll}\toprule
		\textbf{Optics parameters}            & \textbf{Material attributes (FCC aluminum)} \\
		$\NA = 0.72$ mrad                    & $\|\bfb\| = 0.286$ nm\\ 
		$D = 435$ $\mu$m                    & $\nu = 0.334$\\
		$d_1 = 274$ mm                      & $2\theta$ =  $20.73^\circ$ \\
		\textbf{X-ray beam parameters}          &  \textbf{Microscope configuration} \\
 		$\Delta \zeta_v = 0.222$ mrad (top-hat)                 & $\phi = 2.44$ mrad \\
 		$\Delta \zeta_h = 4.25\times 10^{-3}$ mrad (Gaussian)& $\chi = 0.262$ mrad\\
 		$\Delta \zlab = 254.77$ nm (Gaussian) &  $\mu = \theta$ \\
 	    $\Delta E/E = 5.95\times 10^{-5}$ & $\omega = 0$ 
 		
	\end{tabular}

\label{tab:setup}
\end{table}

\section{Bayesian inference for dislocation core position}\label{sec:bayes}

In \Cref{sec:model}, we described the physical model that allows us to simulate the contrast mechanism for an idealized DFXM image of a dislocation at a known position in the sample. For real experimental data collected at synchrotrons, identifying a dislocation's position is difficult for several reasons. Real DFXM images contain measurement noise from multiple sources, with a signal that can be stochastic or deterministic, depending on the mechanism (e.g.,  cosmic rays vs pixel cross-talk) \cite{rossi2006pixel}. As such, when analyzing the DFXM images collected from a time-resolved experiment, we can use intuition from the model \cite{poulsen2020geometrical}, wavelet analysis methods \cite{gonzalez2020methods}, or a combination of methods to estimate the \emph{approximate} position of the dislocation in a given frame. However, our intuition cannot \emph{precisely} locate the dislocation core, nor can it quantify uncertainty in our knowledge of it.

\emph{Bayesian inference} provides a natural framework to refine this initial/imprecise knowledge of the dislocation's position from a DFXM image, and goes a step further by directly quantifying uncertainty in the inferred position, accounting for noise present in the image. Bayesian inference hinges on defining two mathematical objects: a \emph{prior distribution}, which describes our initial knowledge of the dislocation position, and a \emph{likelihood model}, which describes the relationship between the dislocation position and the DFXM images it produces. In other words, the prior distribution describes our initial intuition for the dislocation's position, expressed as a probability distribution. The likelihood model describes the physics that we are using to update our initial understanding of the system. In our work, this physics includes the forward model and experimental noise. Together the likelihood model and prior distribution refine our understanding of the dislocation position, through a distribution termed the \emph{posterior}. The posterior is the prior \emph{updated} or \emph{informed} by our likelihood model and the DFXM image; specifically, it is the distribution of dislocation position \emph{conditioned} on the observed image. Below we discuss the likelihood model and prior distribution we use in this work further.

\subsection{The likelihood model}\label{sec:bayes:likelihood}
In this subsection, we introduce our likelihood model, which integrates our idealized DFXM image simulations from \cref{sec:model} (the forward model $\forward$) with a statistical model of detector measurement noise.
Real DFXM experimental data differ from the idealized simulations due to various sources of random detector noise. To make our position estimation and uncertainty quantification algorithms appropriate for experimental data, we describe the detector using statistical models of measurement noise developed in the literature \cite{snyder1993image,snyder1995compensation}. 
Including both the deterministic forward model and a measurement noise model into the likelihood model lets our algorithms account for both the signal of interest caused directly by dislocations, and the effects of stochastic noise.

Following the work of \cite{snyder1993image}, we describe detector noise based on the types of stochastic processes that occur when light impinges on each pixel of the CMOS camera. Most of these originate from conversion from photons to electrical current. We group the different types of noise processes by their origins and statistical trends. We denote the counts measured by a particular pixel $n \in \{1, \dots, \numpixels \}$ by the random variable $\pixelvalue$. We assume this to be a random variable given by a sum of three mutually independent random variables,
$$
\pixelvalue = \rv{f_n} + \rv{\beta_n} + \rv{\gamma_n},
$$
where $\rv{f_n}$ denotes the counts associated with the feature of interest (i.e., from the material being sampled), $\rv{\beta_n}$ denotes the number of background counts detected (i.e., the background signal), and $\rv{\gamma_n}$ denotes electronic readout noise for that pixel. As noted in \cite{snyder1993image}, both $\rv{f_n}$ and $\rv{\beta_n}$ are accurately modeled using Poisson random variables. Poisson random variables are integer-valued and have equal mean and variance; we denote their means as $\bar{f}_n$ and $\bar{\beta}_n$ respectively.

We take $\bar{f}_n$ to be the deterministic output of the forward model at pixel $n$, i.e., $\bar{f}_{n} = \forward(\pos)_n$, so that the random variable $\rv{f_n}$ encodes our forward model. This implies that in the absence of electronic readout and detector noise, the observed image would be the output of our image simulation. We assume the mean value of the background noise is known and is the same for each pixel, i.e., $\bar{\beta}_n = \bar{\beta}$ for all $n$.

Unlike $\rv{f_n}$ and $\rv{\beta_n}$, we model the electric readout noise measured in each pixel using a Gaussian distribution, $\rv{\gamma_n} \sim \sfN(\bar{\gamma}_n,\sigma^2)$, for convenience. We assume the noise in each pixel to be independent of its neighbors, an assumption that ignores the possibility of pixel \emph{cross-talk}. In this work, we focus on images taken with a weak beam condition, where cross-talk is not significant due to the low integrated pixel intensities. We assume the mean to be $\bar{\gamma}_n = 0$ for all $n$. This value varies greatly with the particular camera used and the gain settings of the camera.

\begin{figure}[h!]
	\centering
		\begin{subfigure}[b]{.27\linewidth}
			\includegraphics[width=\linewidth,trim=3.5cm 1cm 2cm 1cm,clip]{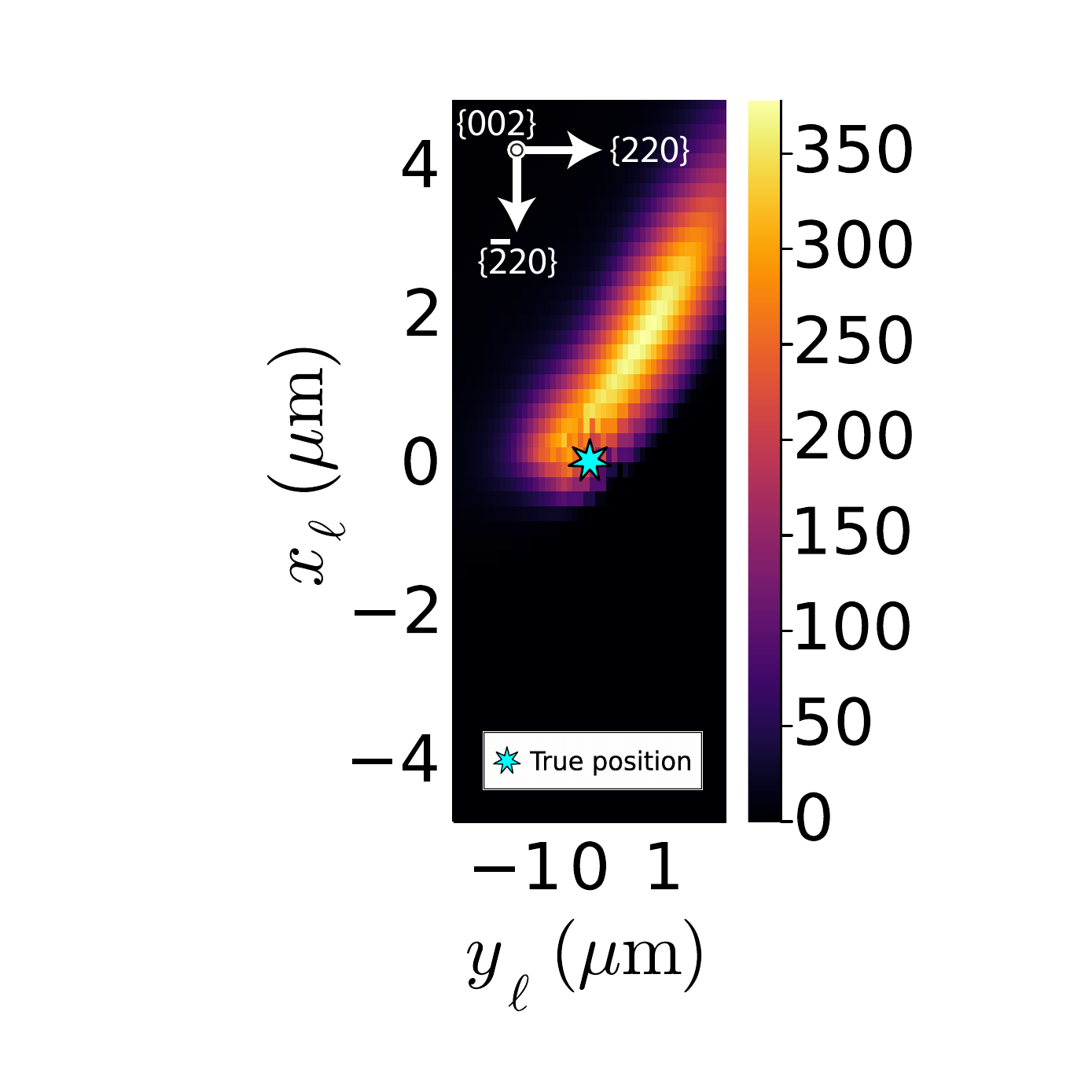}
			\caption{}
			\label{fig:likelihood:true}
		\end{subfigure}
		\begin{subfigure}[b]{.27\linewidth}
			\includegraphics[width=\linewidth,trim=3.5cm 1cm 2cm 1cm,clip]{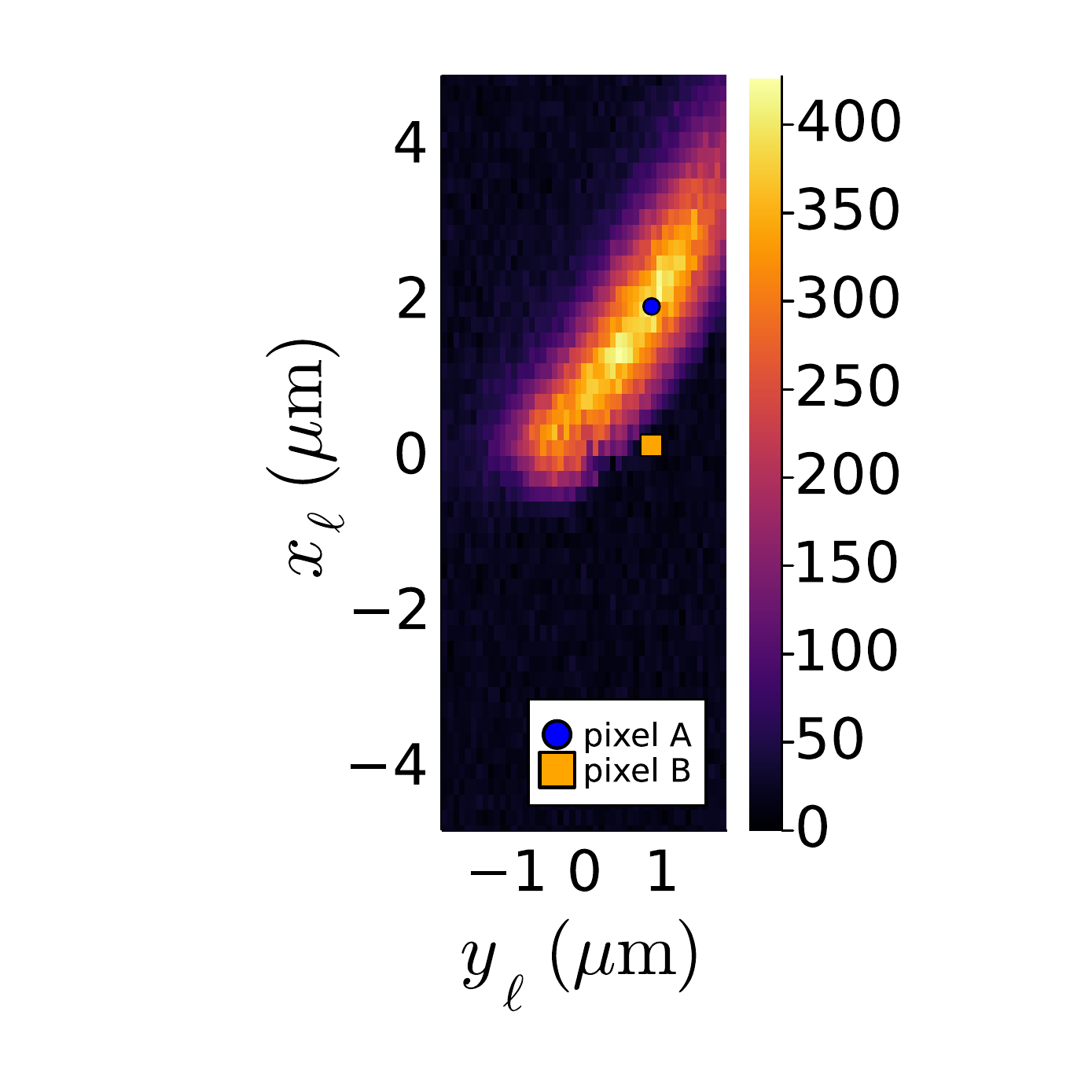}
			\caption{}
			\label{fig:likelihood:noisy}
		\end{subfigure}
		\centering
		\begin{subfigure}[b]{.4\linewidth}
			\includegraphics[width=\linewidth,trim=0cm 1cm 0cm 1cm,clip]{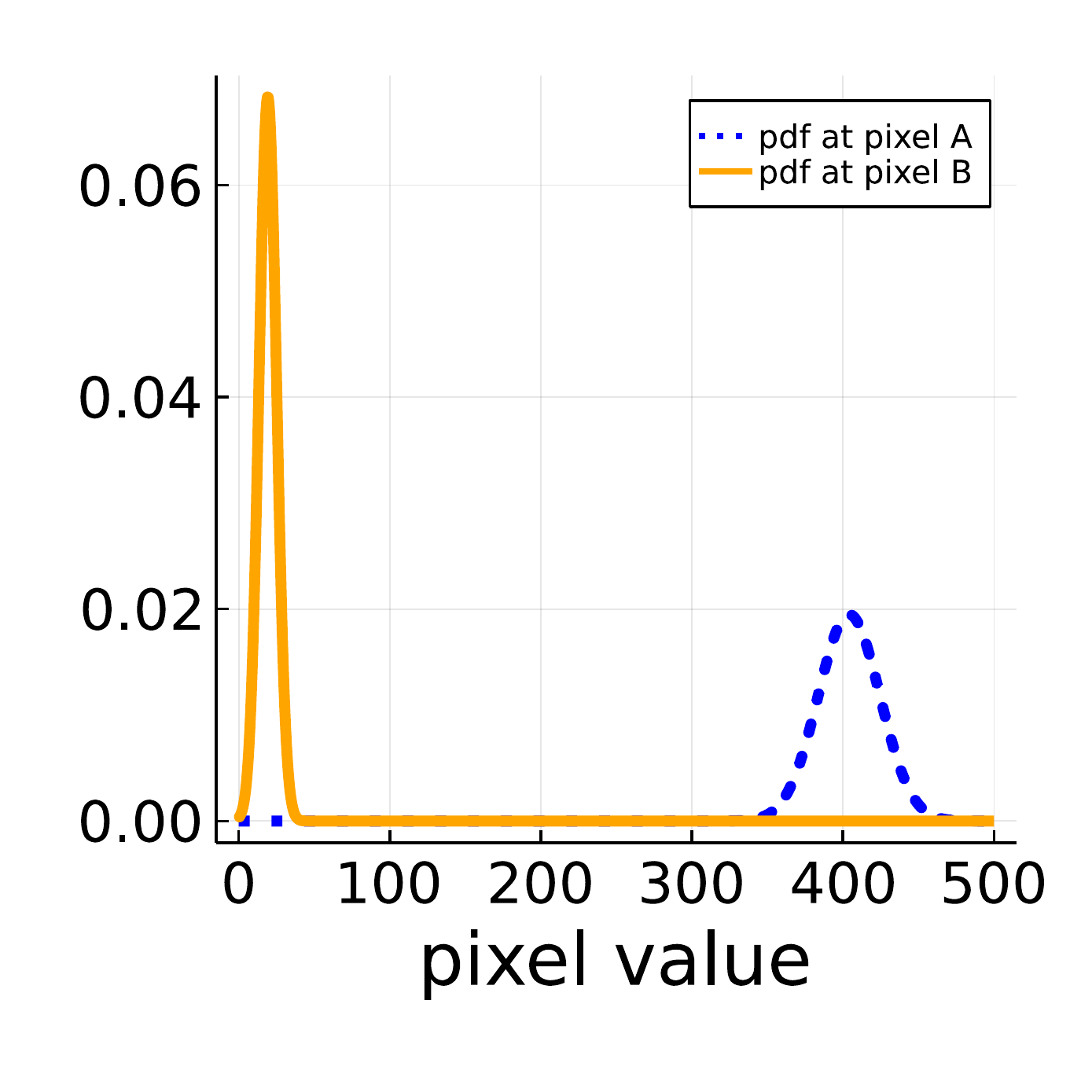}
			\caption{}
			\label{fig:likelihood:pixel1}
		\end{subfigure}	
	\caption{ (a) Noise-free and (b) noisy simulated images for an edge dislocation with Burgers vector $\bfb = \frac{1}{2}[1 \ 0 \ \bar{1}]$ and slip-plane normal vector $\bfn = [1 \ \bar{1} \ 1]$, and distributions at two particular pixels, taking $\bar{\beta}$ and $\sigma^2$ to be $5\%$ and $1\%$ of the maximum noise-free intensity. (c) Intensity pdfs for two pixels showing how a pixel with higher mean intensity also has higher variance.}
	\label{fig:likelihood}
\end{figure}

To obtain a closed-form expression for the probability density function (pdf) of the likelihood model, we make a common simplifying approximation that the sum of Poisson and Gaussian variables can be approximated by a Gaussian distribution,
\begin{equation}
\label{eq:likelihood-dist}
    \pixelvalue \vert \pos \sim \sfN(m_n(\pos), \sigma_n^2(\pos)) \, ,
\end{equation}
with mean intensity measured for each pixel given by
\begin{align*}
    m_n(\pos) &= \bar{f}_n + \bar{\beta}_n + \bar{\gamma}_n \\
    &=\forward(\pos)_n + \bar{\beta}, 
\end{align*}
and variance 
\begin{align*}
    \sigma_n^2(\pos) &=\var{f_n} + \var{\beta_n} + \var{\gamma_n}\\
    &=\forward(\pos)_n + \bar{\beta} + \sigma^2.
\end{align*}
This has the effect that pixels with larger intensities exhibit higher variance, as is seen experimentally. \Cref{fig:likelihood} compares the noise-free idealized simulated DFXM image as described in \cref{sec:model} to a simulated noisy image, a sample drawn from the distribution defined in \cref{eq:likelihood-dist}. We show traces of the intensity pdfs at two particular pixels to demonstrate how the intensity variance depends on its mean intensity.

Under the assumption that the counts measured by different pixels are independent once conditioned on the dislocation position, the conditional pdf of a full image can be expressed as
\begin{equation}
\label{eq:likelihood-function}
    \pi(\obs \vert \pos)  =  \mathcal{N} \left (\obs; \bfm(\pos), \boldsymbol{\Sigma}(\pos) \right ),
\end{equation}

where $\bfm(\pos) = \left[ m_1(\pos), \dots, m_{\numpixels}(\pos) \right]$ and $\boldsymbol{\Sigma}(\pos) = \diag\left[\sigma^2_1(\pos), \dots, \sigma^2_{\numpixels}(\pos) \right]$. In \cref{sec:bayes:posterior}, we will set the variable $\obs$ to the image being evaluated, $\trueobs$. The \emph{likelihood function} is then obtained by viewing $\pi(\trueobs \vert \pos)$ as function of $\pos$.

\subsection{The prior distribution}\label{sec:bayes:prior}
\begin{figure}[t]
	\centering
	\begin{minipage}{.24\linewidth}
		\includegraphics[width=\linewidth,trim=3.5cm 1cm 2cm 1cm,clip]{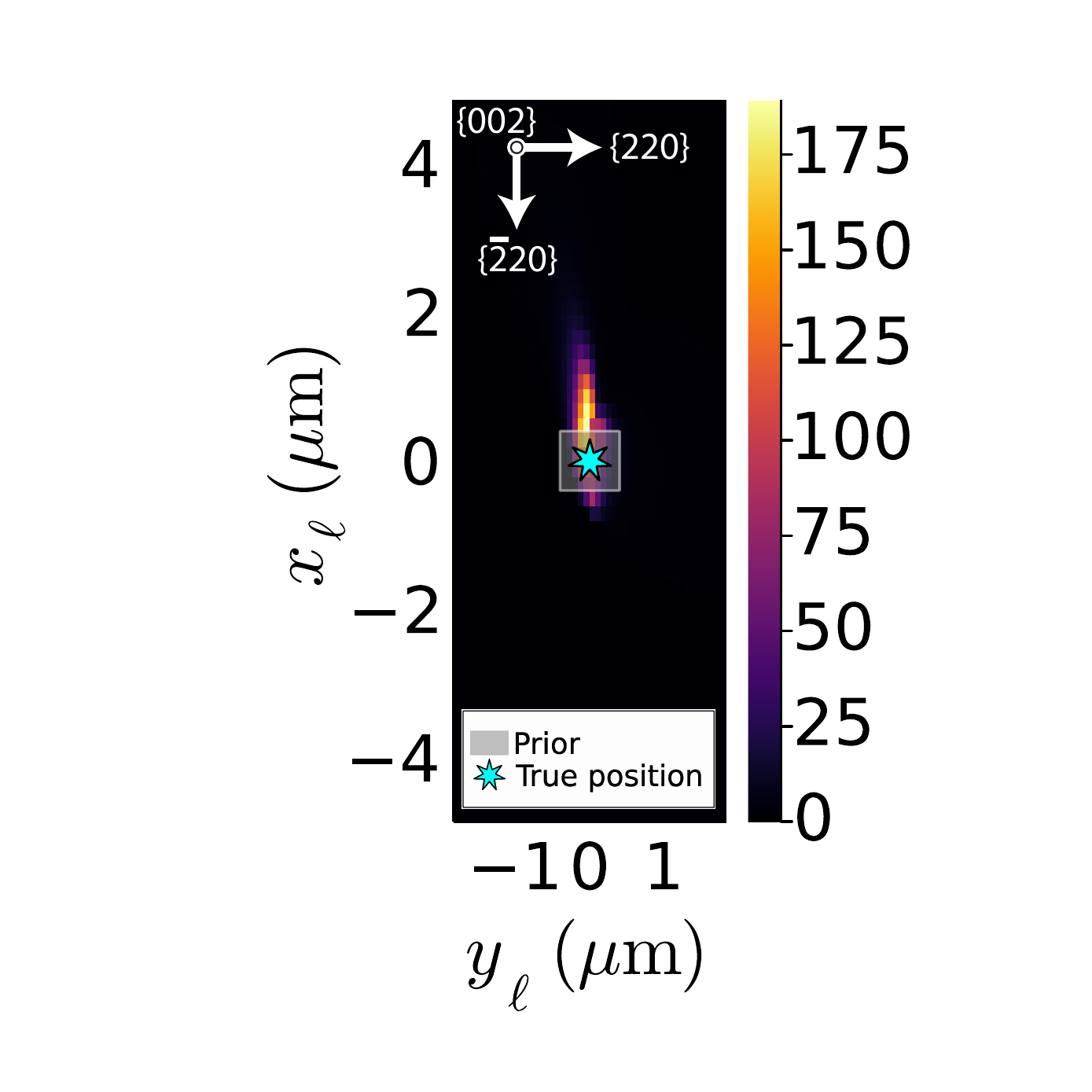}
	\end{minipage}
	\begin{minipage}{.24\linewidth}
		\includegraphics[width=\linewidth,trim=3.5cm 1cm 2cm 1cm,clip]{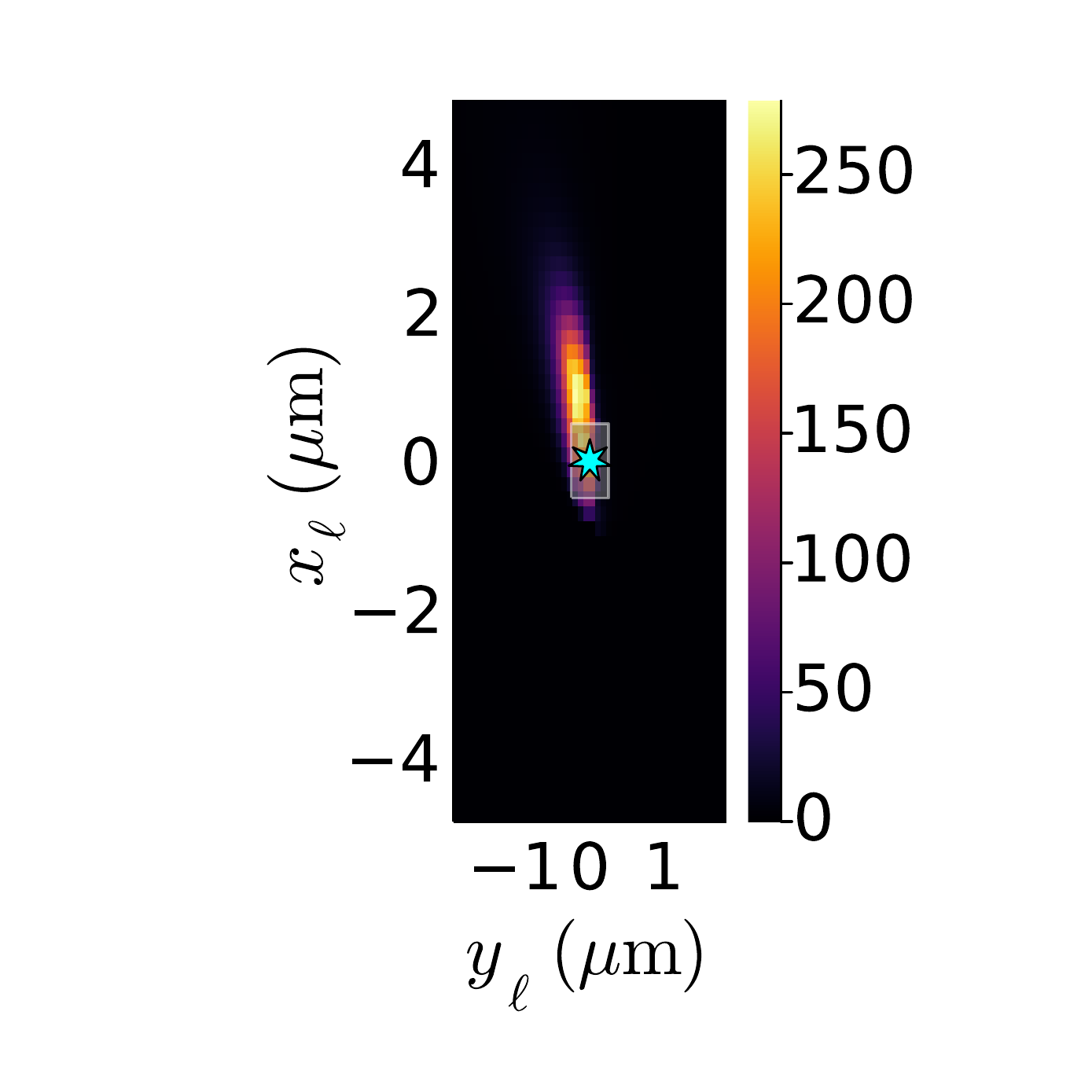}
	\end{minipage}
	\begin{minipage}{.24\linewidth}
		\includegraphics[width=\linewidth,trim=3.5cm 1cm 2cm 1cm,clip]{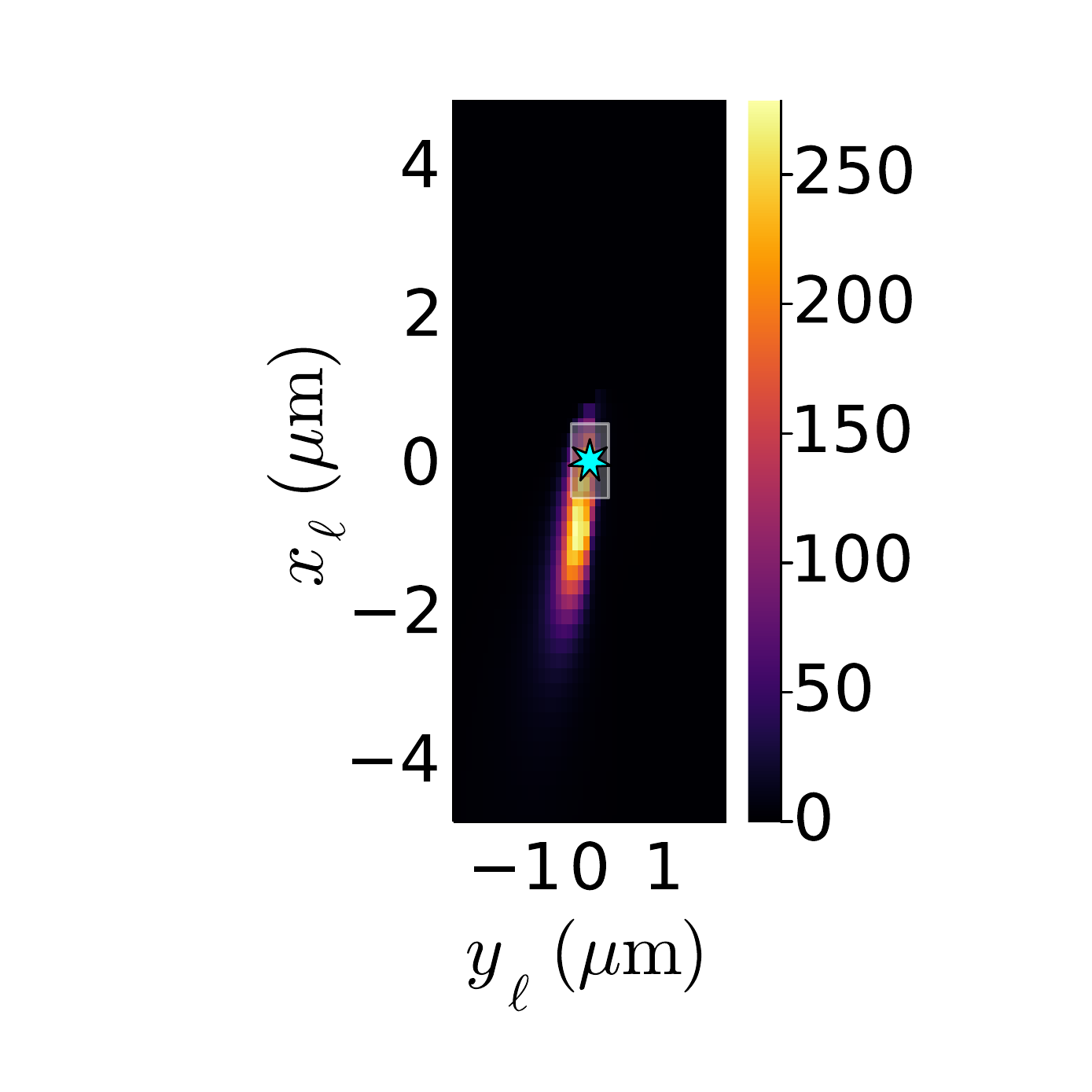}
	\end{minipage}
	\begin{minipage}{.24\linewidth}
		\includegraphics[width=\linewidth,trim=3.5cm 1cm 2cm 1cm,clip]{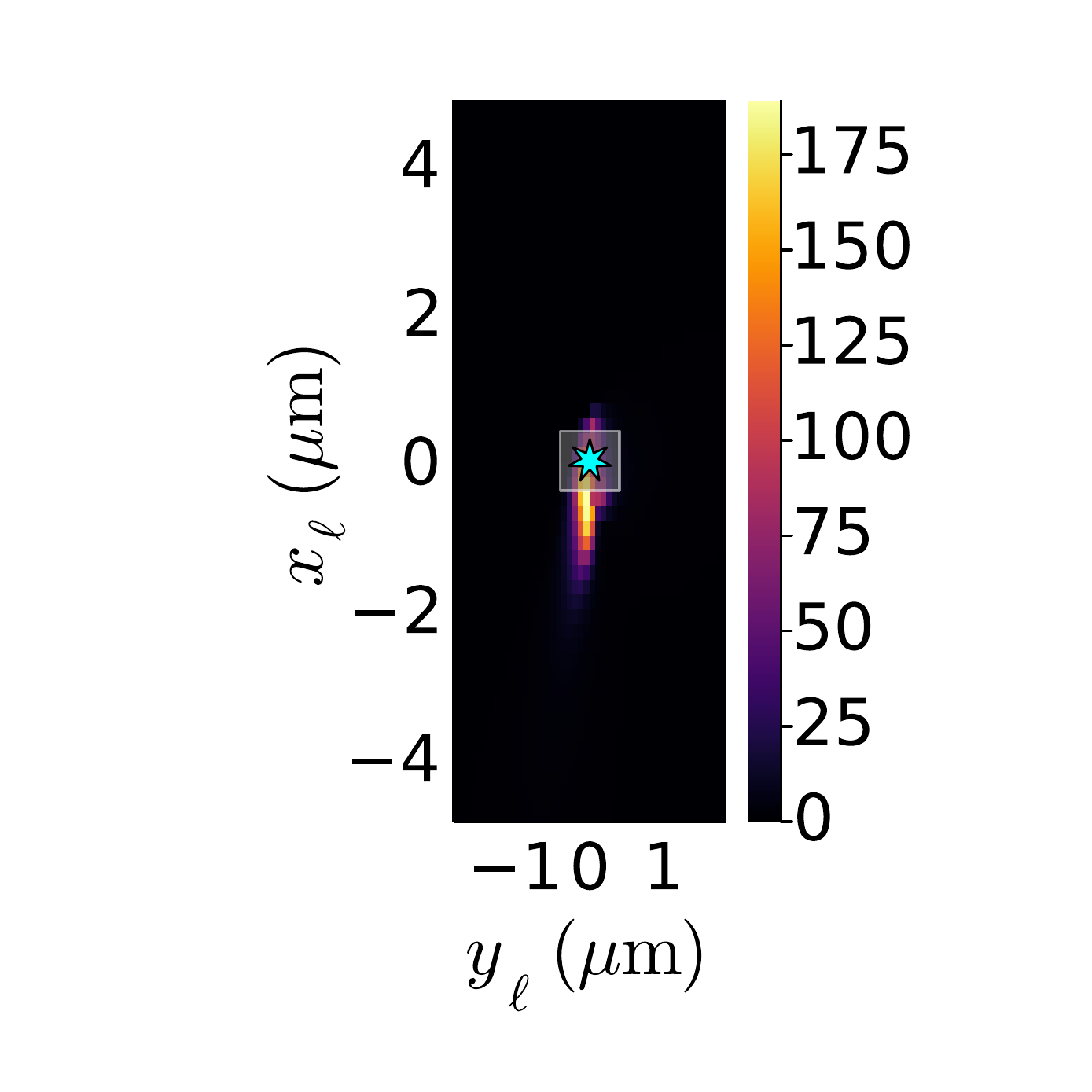}
	\end{minipage}
	\begin{minipage}{.24\linewidth}
		\includegraphics[width=\linewidth,trim=3.5cm 1cm 2cm 1cm,clip]{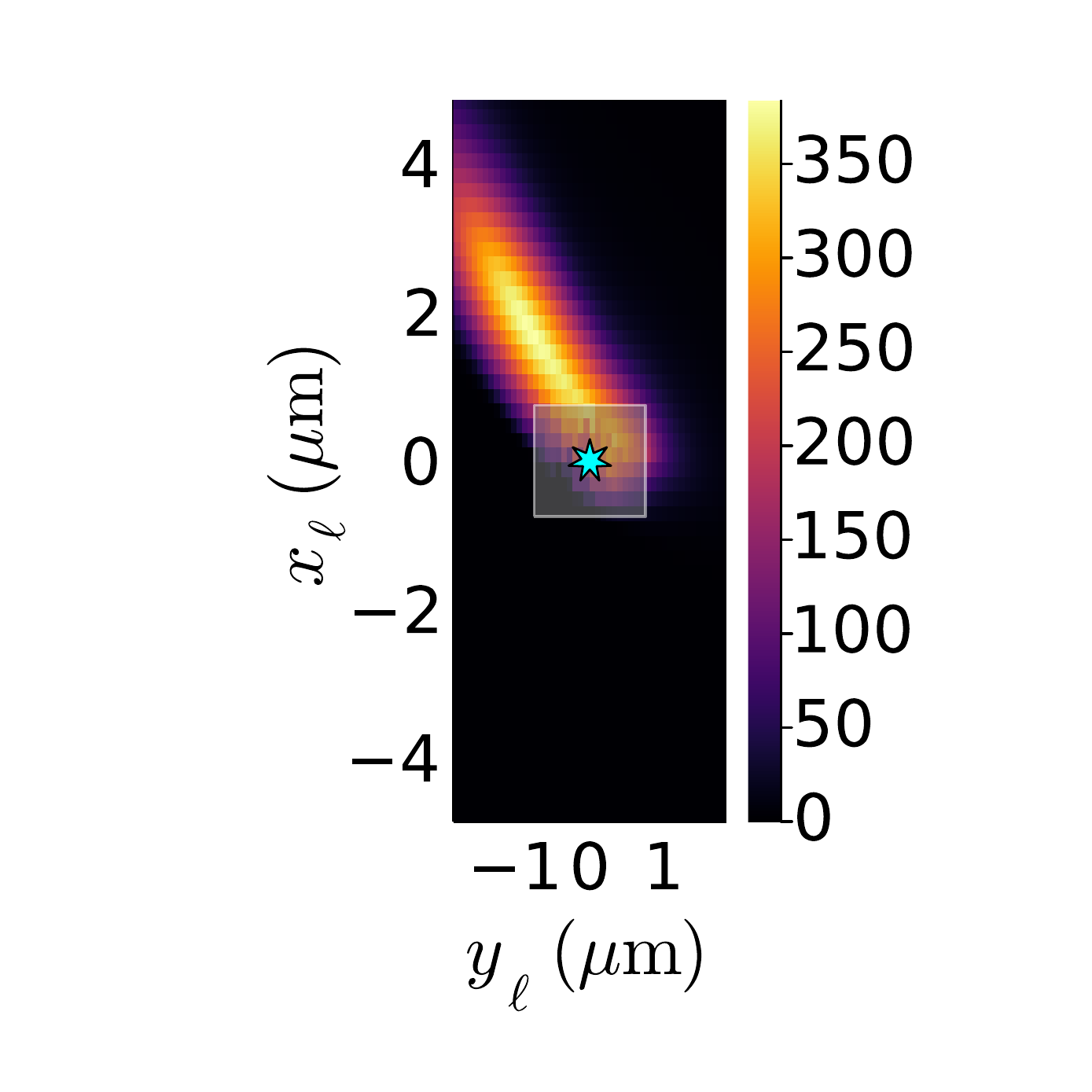}
	\end{minipage}
	\begin{minipage}{.24\linewidth}
		\includegraphics[width=\linewidth,trim=3.5cm 1cm 2cm 1cm,clip]{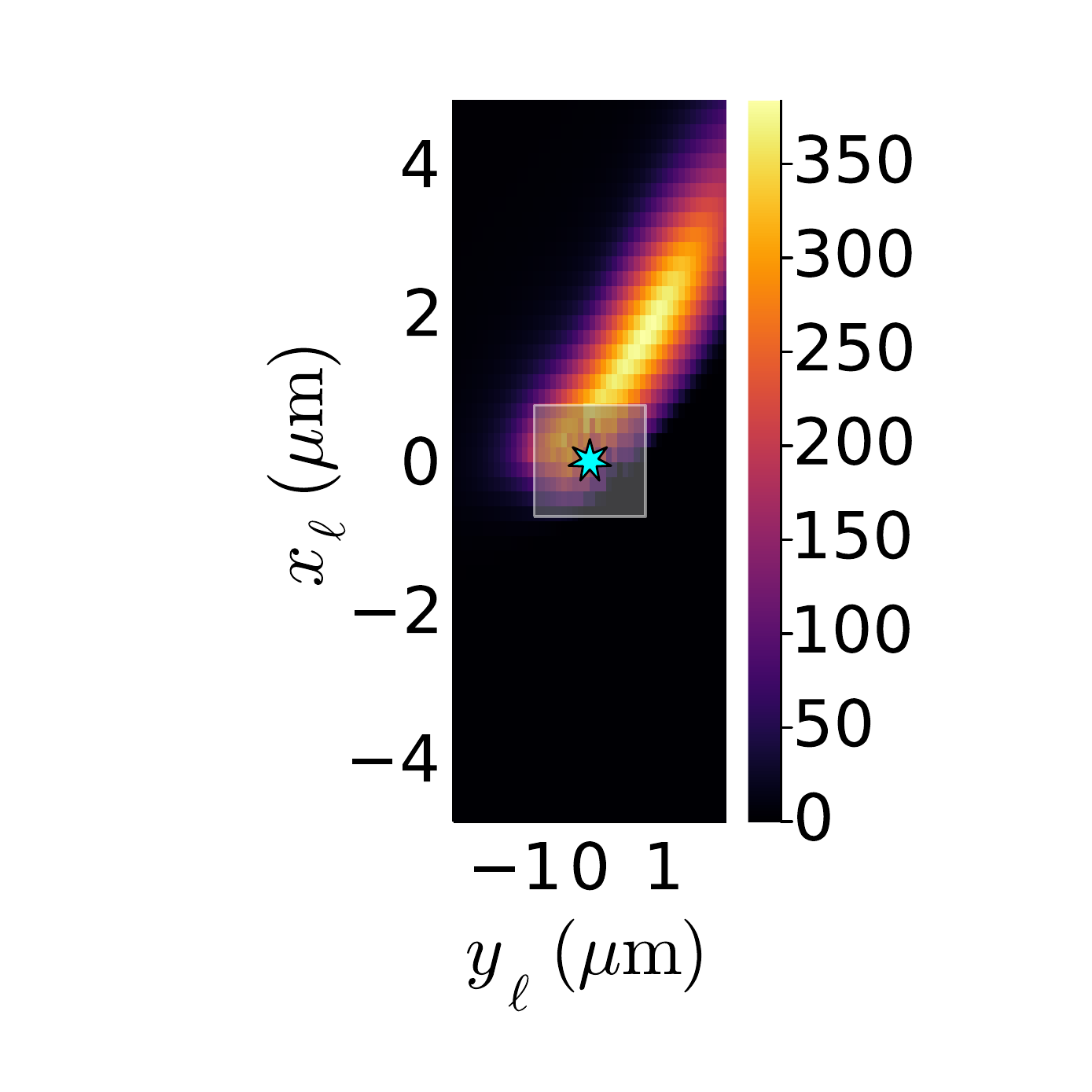}
	\end{minipage}
	\begin{minipage}{.24\linewidth}
		\includegraphics[width=\linewidth,trim=3.5cm 1cm 2cm 1cm,clip]{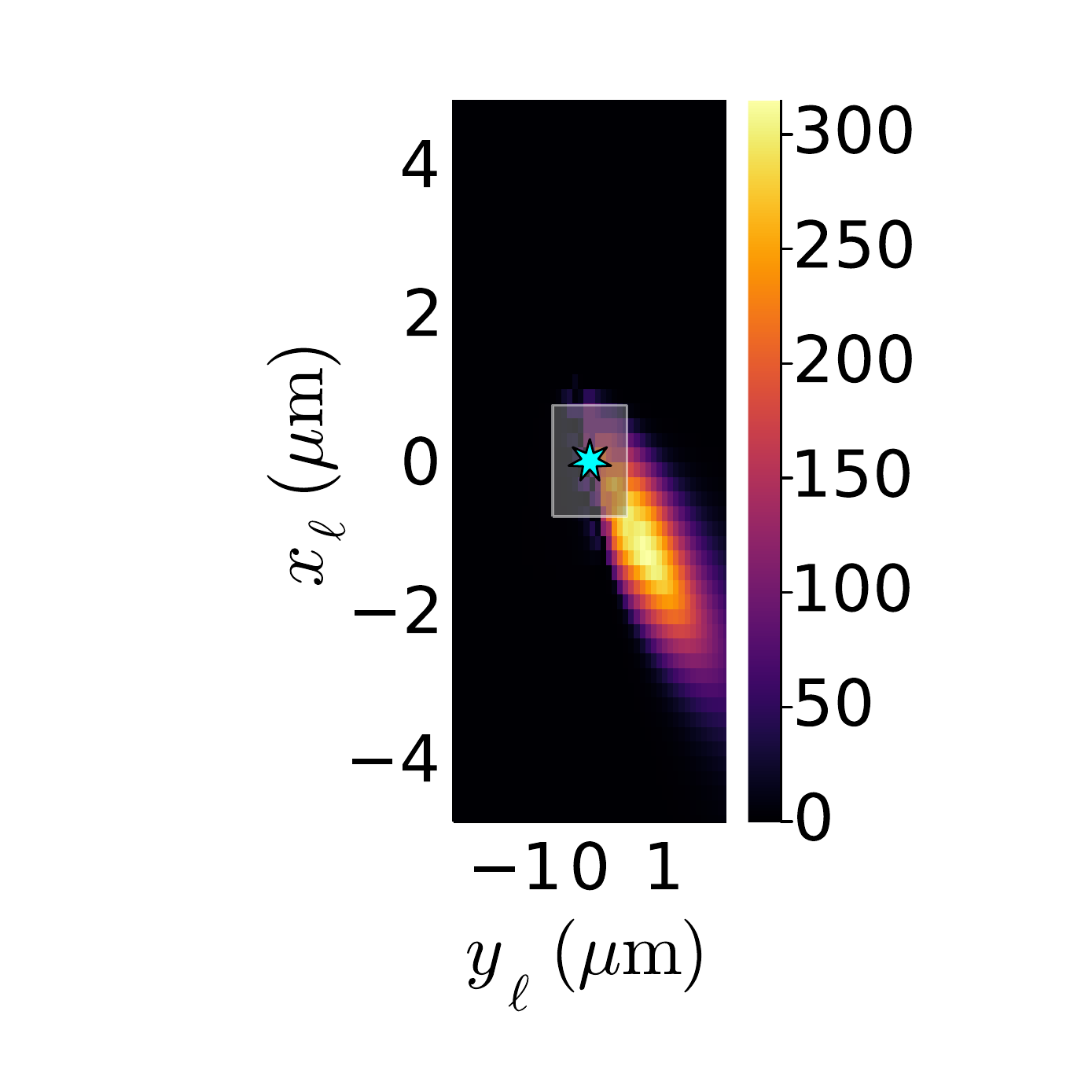}
	\end{minipage}
	\begin{minipage}{.24\linewidth}
		\includegraphics[width=\linewidth,trim=3.5cm 1cm 2cm 1cm,clip]{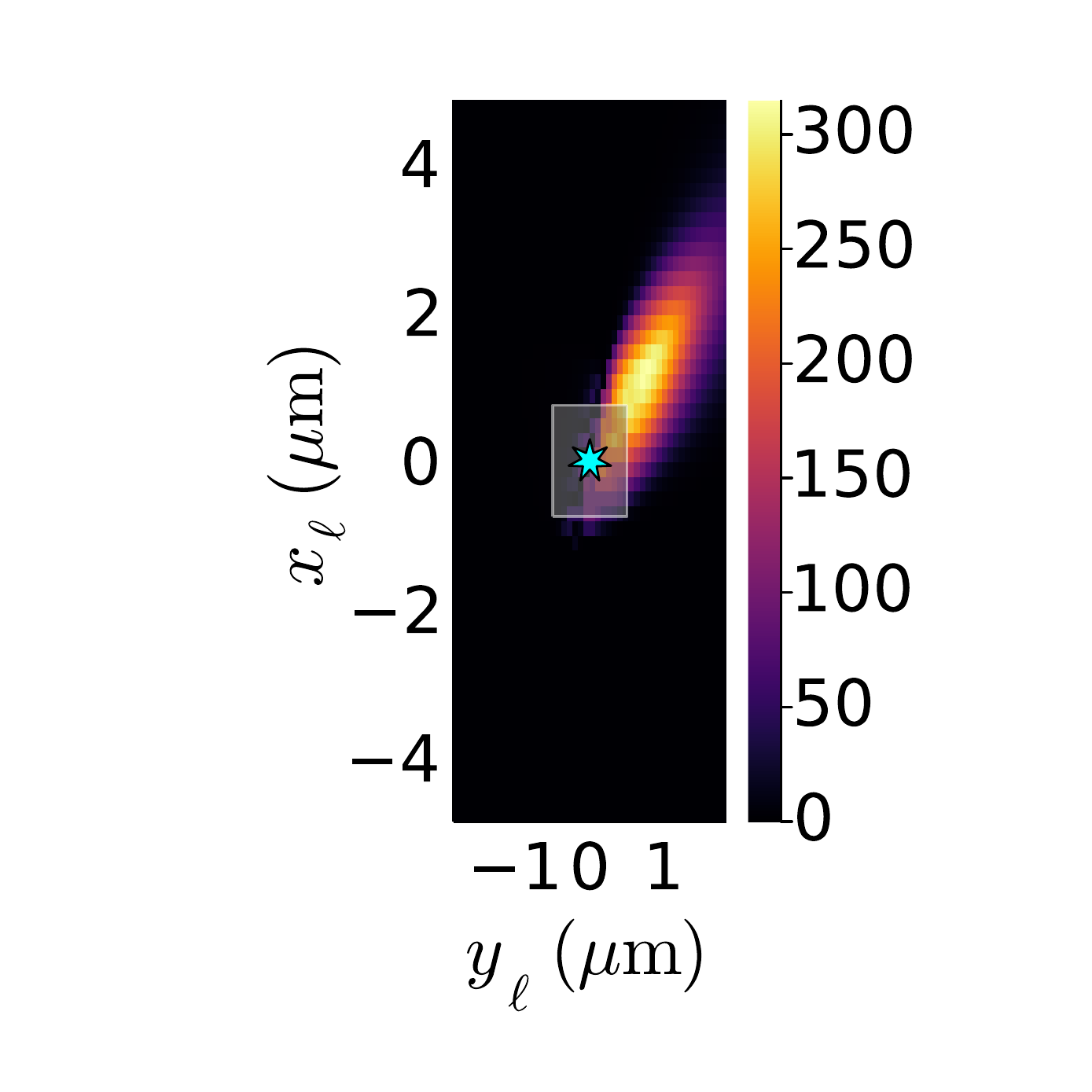}
	\end{minipage}
	\begin{minipage}{.24\linewidth}
		\includegraphics[width=\linewidth,trim=3.5cm 1cm 2cm 1cm,clip]{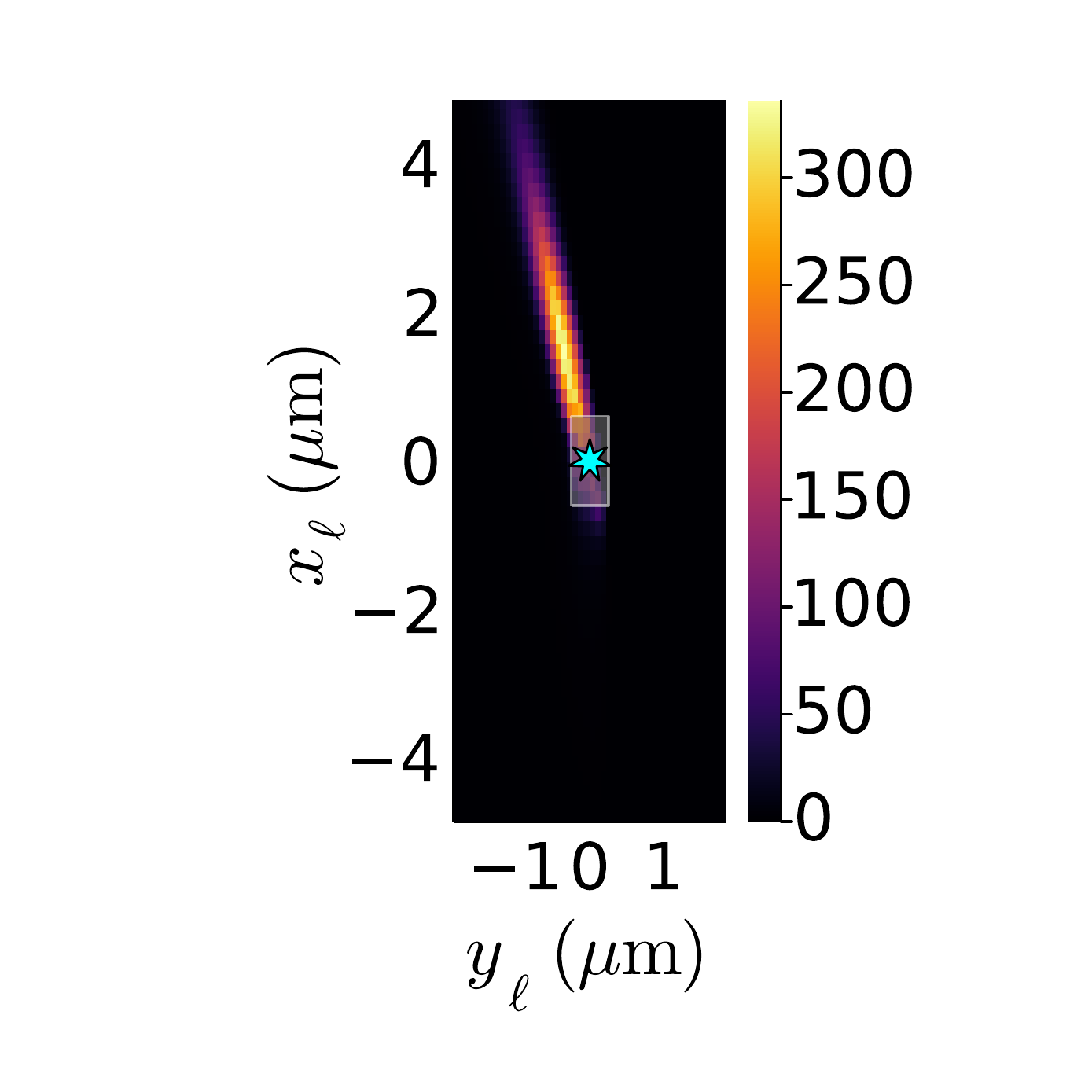}
	\end{minipage}
	\begin{minipage}{.24\linewidth}
		\includegraphics[width=\linewidth,trim=3.5cm 1cm 2cm 1cm,clip]{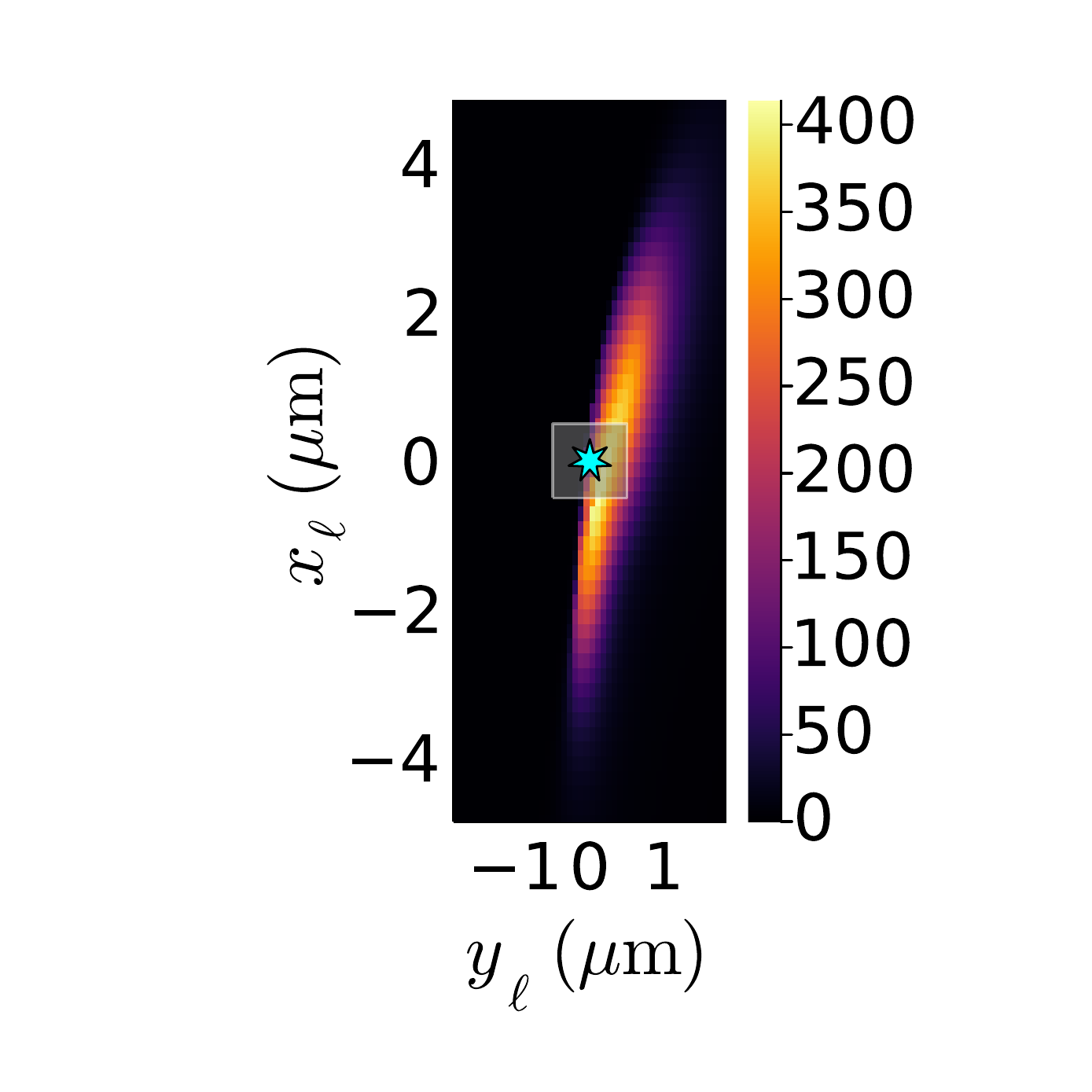}
	\end{minipage}
	\begin{minipage}{.24\linewidth}
		\includegraphics[width=\linewidth,trim=3.5cm 1cm 2cm 1cm,clip]{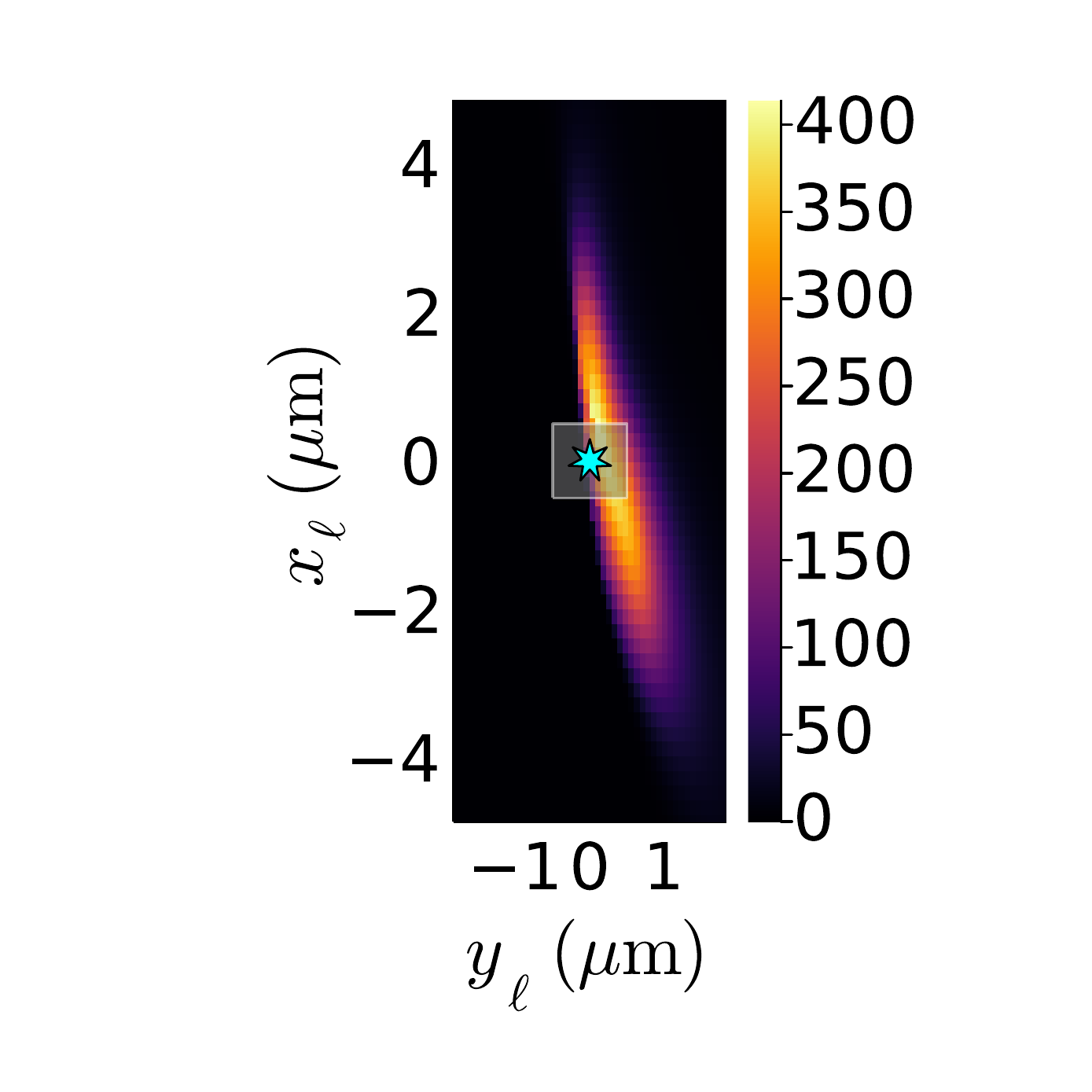}
	\end{minipage}
	\begin{minipage}{.24\linewidth}
		\includegraphics[width=\linewidth,trim=3.5cm 1cm 2cm 1cm,clip]{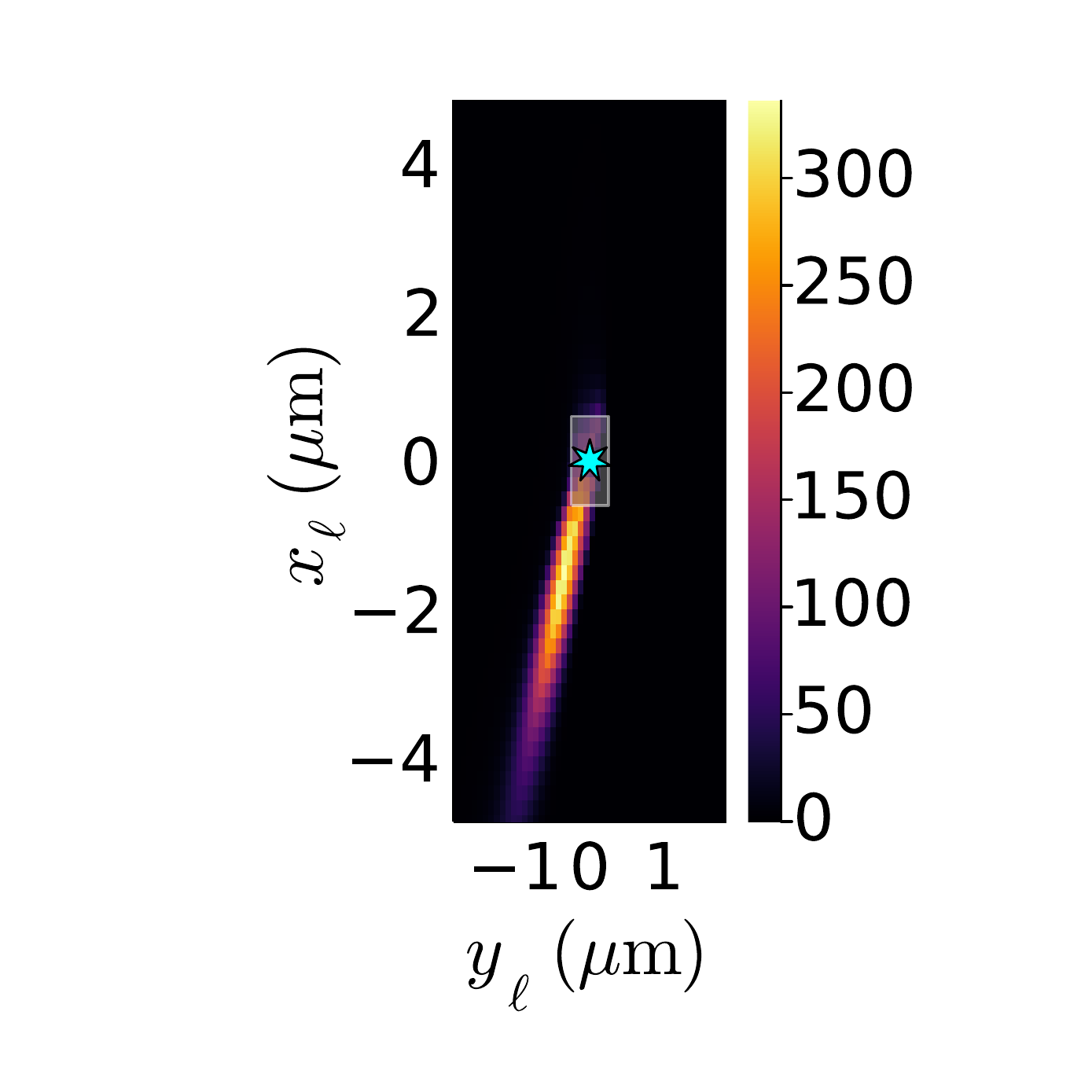}
	\end{minipage}

	\caption{Simulated DFXM images for the twelve possible characters known
for edge dislocations in FCC aluminum using the parameters listed in \cref{tab:setup} with the support of the prior distribution highlighted with a gray box.}
	\label{fig:prior}
\end{figure}

The \emph{prior distribution} (or \emph{the prior}) is a probability distribution that encodes our knowledge of the dislocation position, $\pos$, before evaluating the DFXM image.
For this work, we assume the intuition afforded by the forward model will be sufficient to discern the dislocation character based on its image features. We found this assumption to be valid for the case of isolated dislocations. 
Thus, for a known dislocation character, we describe our initial ``lower confidence'' estimate of the dislocation core position with a box of uniform probability over a region representing a reasonable margin of uncertainty, i.e., based on the feature size, and we assign zero probability outside of this region. This region is referred to as the \emph{support} of the prior.
In our simulated study, since we have a well defined ground truth, for the dislocation position we define the prior's support for each of the $12$ edge dislocations that appear in FCC lattices differently. To show our assumptions for user uncertainty, we plot the support of the prior on top of the noise-free simulated image for each dislocation character in \Cref{fig:prior}.
We note that in general, the choice of the prior distribution will depend highly on the specific experiment and application of interest. We discuss possible modifications to our prior in \cref{sec:discussion}.

\subsection{The posterior distribution}
\label{sec:bayes:posterior}

The \emph{posterior distribution} describes our knowledge of the dislocation position after evaluating an image with our likelihood model. Let $\trueobs$ denote the hypothetical observed image collected by a DFXM experiment. Let $\pi(\pos)$ denote the pdf of the prior, and let $\pi(\obs \vert \pos)$ denote the likelihood function defined in \cref{eq:likelihood-function}. The prior distribution and likelihood model then define the pdf of the posterior distribution $\pi(\pos\vert\trueobs)$ through Bayes' rule
$$
\pi(\pos\vert\trueobs) = \frac{\pi(\trueobs\vert\pos)\pi(\pos)}{Z},
$$
where $Z$ is a normalization constant. The posterior captures our updated understanding of which dislocation positions are more and less probable given the observed image, weighed by our initial prior beliefs. Information that can be extracted from the posterior  thus includes \emph{point estimates} of the dislocation position, as well as complete characterizations of uncertainty in the inferred position. 

\section{Algorithms for estimating and quantifying uncertainty in dislocation core positions}
In our Bayesian framework, the tasks of position estimation and uncertainty quantification can be accomplished by characterizing the posterior distribution. This is commonly done by computing summary statistics of the posterior distribution. For the task of position estimation,  we can use the posterior mean or mode, and we can measure our confidence in those predictions using probabilistic measures of spread like the marginal posterior variances or the entire posterior covariance matrix. To this end, we propose three algorithms for estimation and uncertainty quantification:
\begin{itemize}
	\item Estimation via the \textit{maximum a posteriori} (MAP) point.
	
	\item Approximation of the posterior using the Laplace approximation.
	
	\item Computing summary statistics of the posterior using a Markov chain Monte Carlo (MCMC) method.
\end{itemize}
These approaches have increasing levels of fidelity to the posterior, at the cost of increasing amounts of required computational time. In the following subsections we introduction each option and discuss its strengths and weaknesses. We provide the numerical implementation of each method and experiment scripts at \githublink. 

\subsection{MAP point estimation}
Our primary goal is to estimate the position of each dislocation based on a hypothetical observed image measured by DFXM. One natural choice for our estimate is the MAP point,\footnote{In general the MAP point is not unique, as there may be many points where the posterior pdf achieves its maximum.} the position that maximizes the posterior pdf,
$$
\mappos = \argmax_\pos \pi(\pos\vert\trueobs).
$$
Note that Bayes' rule allows us to compute the posterior pdf up to an unknown normalizing constant, $Z$. As such, we define the unnormalized version of the posterior pdf as $\widetilde{\pi}(\pos\vert\trueobs) =  \pi(\trueobs\vert\pos)\pi(\pos)$, and recast the problem as
$$
\mappos = \argmax_\pos \widetilde{\pi}(\pos\vert\trueobs).
$$
For computational convenience, we consider the equivalent minimization problem of the negative posterior log-pdf
$$
\mappos = \argmin_\pos \left\{ -\log\widetilde{\pi}(\pos\vert\trueobs)\right\} = \argmin_\pos \left\{-\log\pi(\pos)-\log\pi(\trueobs\vert\pos)\right\},
$$
which results in a more numerically stable optimization problem. We used the \verb|Optim| Julia package to solve for the MAP estimate using a gradient descent method \cite{mogensen2018optim}.

\subsection{Laplace approximation}
The Laplace approximation \cite{raftery1996approximate} approximates the posterior with a Gaussian distribution centered at the MAP point, with covariance matrix set to match the local curvature of the posterior; specifically, the Laplace covariance is the negative Hessian of the log posterior pdf at the MAP point. 
If the posterior distribution were truly Gaussian, this approximation would match the posterior exactly. The covariance matrix of the Laplace approximation enables approximate uncertainty quantification, in the sense that it can closely resemble the true posterior covariance, particularly when the posterior is unimodal.

In our examples we plot $2$-standard deviation credibility ellipses ($2\sigma$ ellipses) to visualize this uncertainty. We note that while the posterior distributions generated in our numerical examples do have non-Gaussian features, the Laplace approximation still correctly captures the directions of greatest uncertainty. 

\subsection{MCMC sampling}

Finally, we describe a method to most precisely characterize the posterior distribution, MCMC sampling \cite{brooks2011handbook}. MCMC methods generate samples according the posterior distribution by using a Markov chain designed to have the posterior as its unique stationary measure, and to converge to this measure from any starting point. In other words, MCMC samples are asymptotically \emph{exact}. Given a set of MCMC samples, $\{\pos_i\}_i$, one can compute sample estimates of any posterior moment or expectation. For example, the sample mean can be used as an estimate for the dislocation position, and the sample variance provides a measure of uncertainty in the position. The MCMC samples also show us if the posterior has non-isotropic, multi-modal, or generally non-Gaussian features. In our numerical experiments, we used a simple adaptive Metropolis method \cite{haario2001adaptive, vihola2014ergonomic}.

\section{Results}\label{sec:results}

\begin{figure}[h!]
    \centering
    \includegraphics[width=\linewidth]{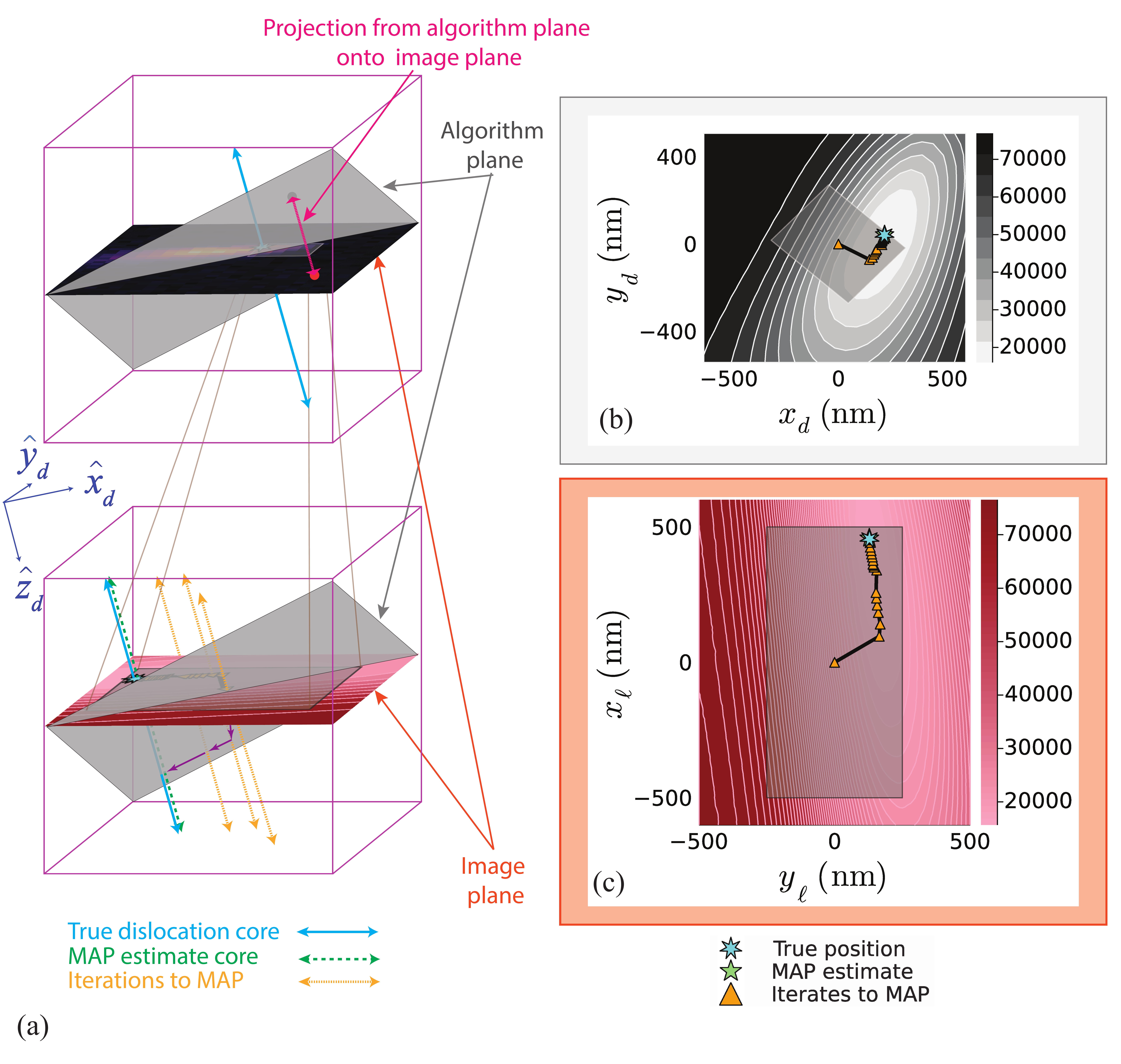}
    \caption{ (a) The evaluation image and MAP optimization results shown in the image plane in respect to the algorithm plane. The true dislocation position, MAP estimate and several iterates are shown as dislocation cores aligned with the known line vector. (b/c) Results of the MAP optimization shown in the algorithm and image plane respectively. We show the true contours of posterior log-pdf to help visualize the behavior of the optimization method.}
    \label{fig:3d-diagram-and-map}
\end{figure}

In this section we present the performance of our estimation and uncertainty quantification algorithms for images of single edge dislocations in single-crystal aluminum. As described in \cite[Chapter~4.3]{hull2011introduction}, the displacement gradient tensor field $\nabla \bm{u}(x,y,z)$ induced by a theoretical infinite length dislocation is invariant along the direction of the dislocation line vector. Indeed, in this model system, $\nabla \bm{u}$ can be expressed analytically at all 3D points in an arbitrary coordinate system, or described as a 2D function along the coordinate system defined by the Burgers vector $\xdisvec$ and slip plane normal $\ydisvec$. Since we assume a known character and therefore line vector for the dislocation, its core position can be represented in 2D by a point defined in the plane orthogonal to the line vector. We define this plane to contain the original origin from the lab frame and call this the \emph{algorithm plane}. 
Expressing our algorithms in this coordinate system captures our intuition that we should have infinite uncertainty in the $\zdisvec$ direction for an idealized infinite length dislocation. We emphasize the fictitious nature of the algorithm plane, as we can still truthfully represent finite-length dislocations using this system, so long as they behave this way within the span of the observation plane.
\Cref{fig:3d-diagram-and-map}a. visualizes the algorithm plane along with the observation plane and dislocation coordinate system for a dislocation with Burgers vector $\bfb = \frac{1}{2}[1 \ 1 \ 0]$, and slip plane normal $\bfn = [1 \ \bar{1} \ 1]$. One may interpret inference results in the lab frame by projecting points in the algorithm plane onto the observation plane in the direction of the line vector as show in \cref{fig:3d-diagram-and-map}. If another coordinate frame is of interest, one can project points from the algorithm plane into that coordinate frame similarly. 

In order to evaluate the accuracy of our estimation methods, we require evaluation images with a known true dislocation position (i.e., the ``ground truth''). To achieve this, we focus this first study on synthetic ``experimental'' images that we generate with the likelihood. Thus, for the true position $\truepos$, we draw a sample $\trueobs$ from the likelihood distribution defined by \cref{eq:likelihood-dist}. We set the magnitude of the corrupting detector-noise processes by setting $\bar{\beta}$ and $\sigma^2$ to be $5\%$ and $1\%$ of the maximum noise-free intensity; this value matches the noise we have seen in our experiments, and is thus representative of real experiments \cite{Yildirim2022,jakobsen2019mapping}

\begin{figure}[h!]
\centering
\includegraphics[width=\linewidth]{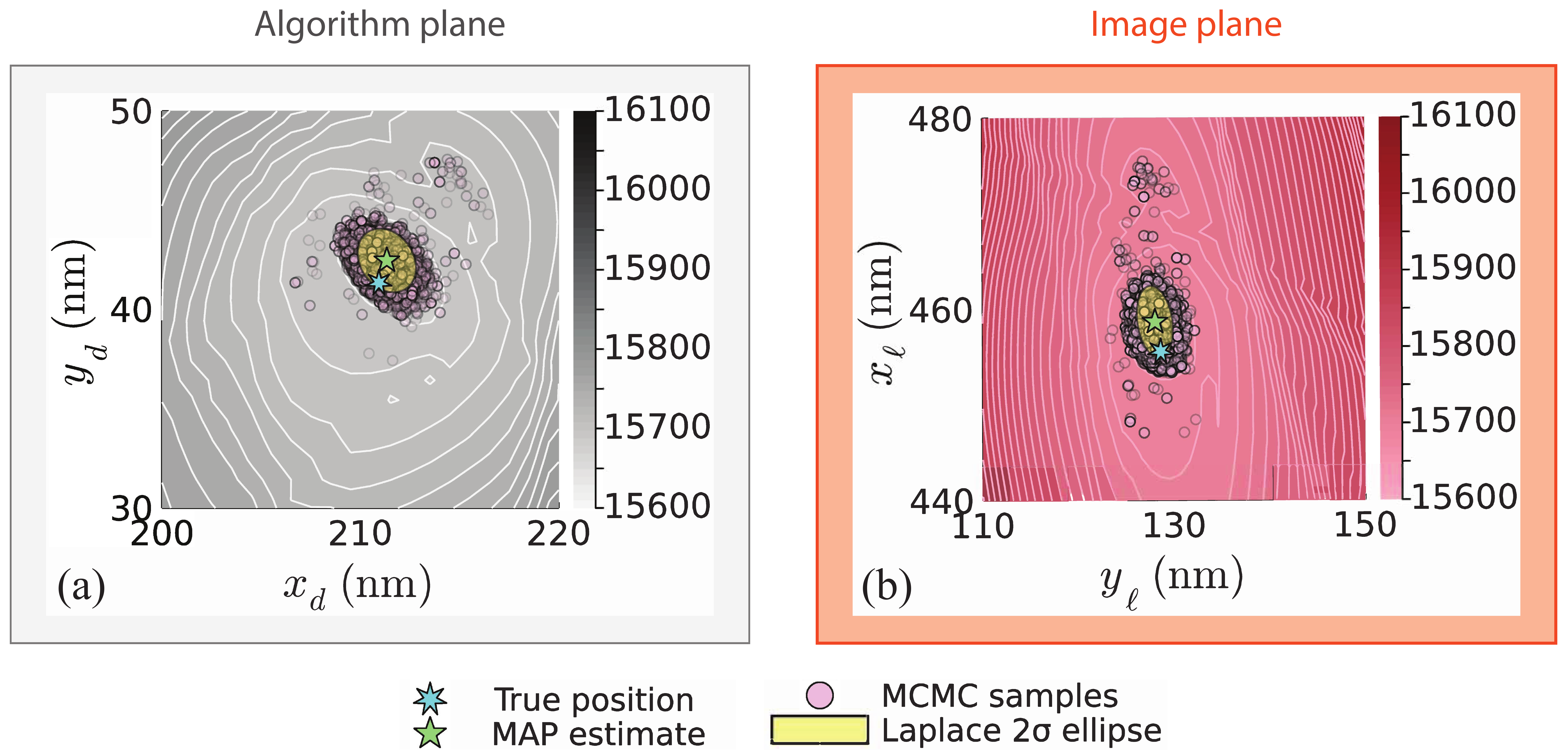}
	\caption{Visualizations of the results of the Laplace approximation and MCMC sampling algorithms (a) in the algorithm plane (b) in the image plane}
	\label{fig:uq_results}
\end{figure}

We now show the outputs of each estimation algorithm introduced in \cref{sec:bayes} for the edge dislocation with Burgers vector $\bfb = \frac{1}{2}[1 \ 1 \ 0]$, and slip plane normal $\bfn = [1 \ \bar{1} \ 1]$, where the true dislocation position was generated randomly within the support of the prior.
\Cref{fig:3d-diagram-and-map}b and \Cref{fig:3d-diagram-and-map}c  show optimization iterations, starting at the origin and ending at the MAP point $\mappos$ in the algorithm and image plane respectively. In this example the MAP point is approximately $3$ nm from the true dislocation position. We note that, given the $75$ nm pixel size, this demonstrates a error less than $10\times$ smaller than the pixels on the detector.
\Cref{fig:uq_results} shows the Laplace approximation and MCMC samples for this example in both the algorithm and image planes. We see that the Laplace approximation closely matches the MCMC samples in both frames in terms of shape, but slightly under-reports the uncertainty in the position, indicated by many MCMC samples lying outside of the $2\sigma$ ellipses. The MCMC samples also show us that the true posterior does indeed have some anisotropic characteristics. In particular, both algorithms show that the direction of highest uncertainty is closely aligned with $\xlab$ direction in the image plane.

\begin{figure}[h!]
	\centering
	\begin{subfigure}[b]{.45\textwidth}
	\includegraphics[width=\linewidth]{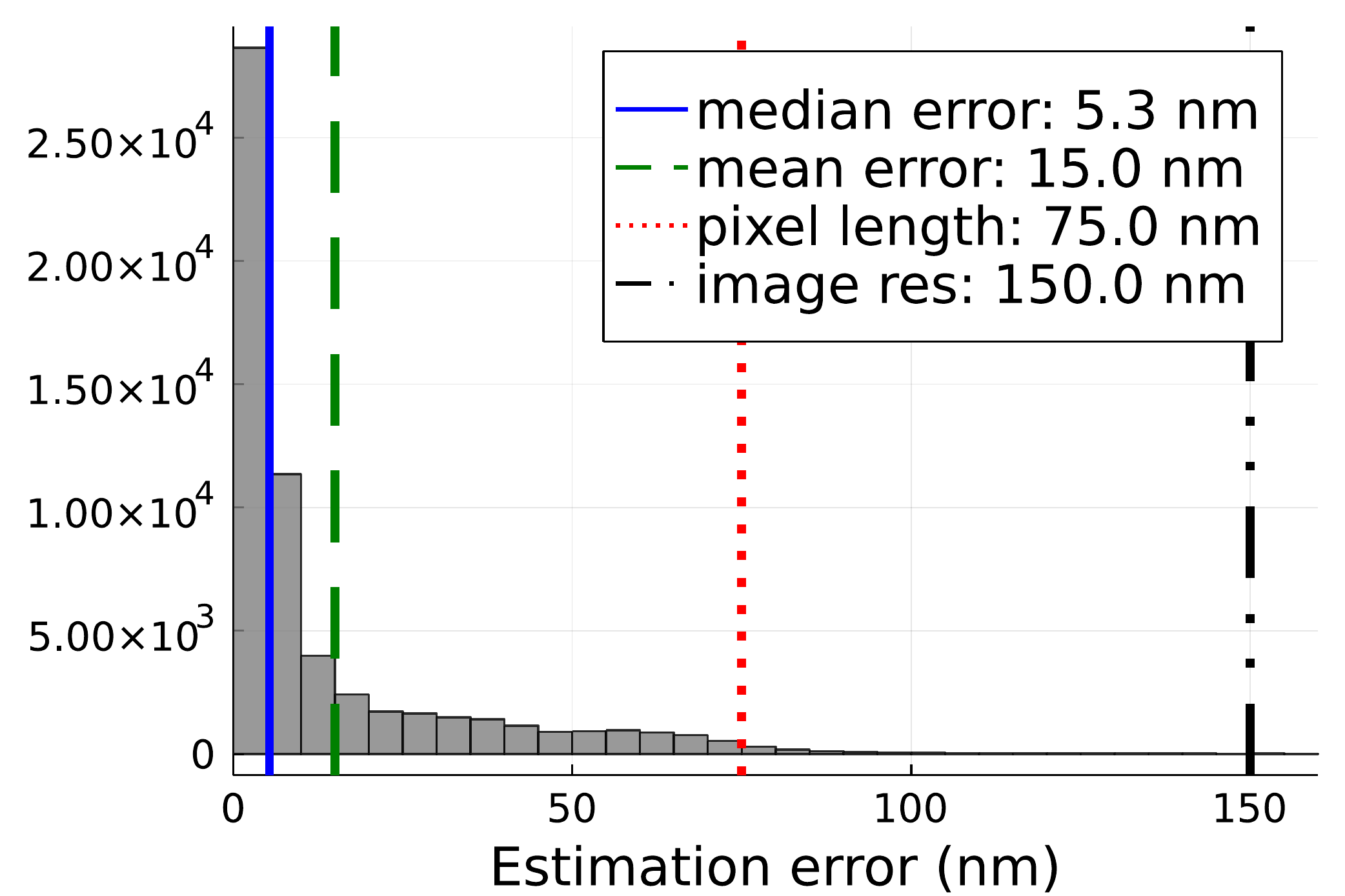}
	\caption{The low noise case}
	   	\label{fig:error-histos:low}
	\end{subfigure}
	\begin{subfigure}[b]{.45\textwidth}
        \includegraphics[width=\linewidth]{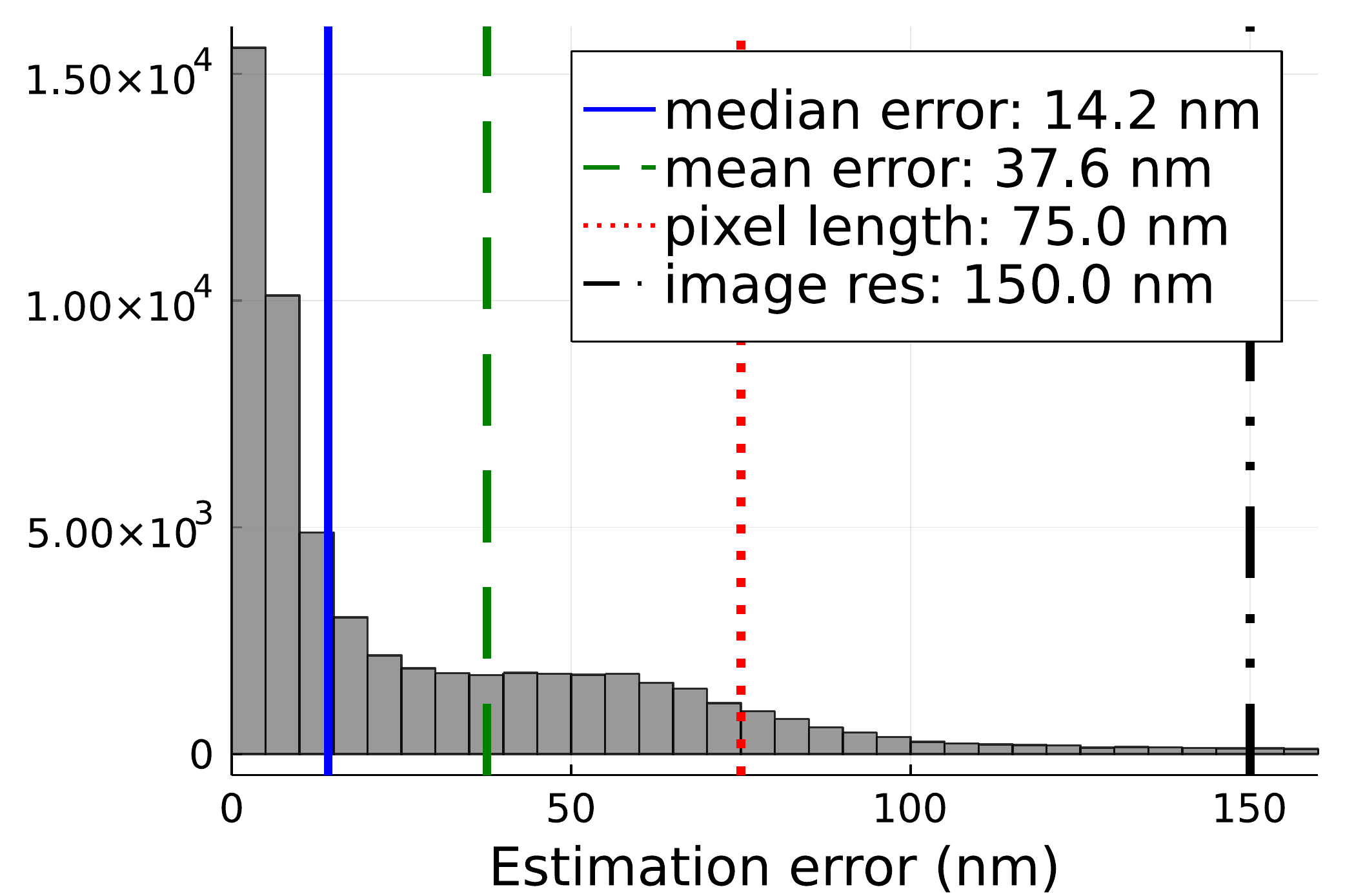}
   \caption{The high noise case}
   	\label{fig:error-histos:high}
	\end{subfigure}
	
	\caption{Histograms of error between $\truepos$ and $\mappos$ for the (a) low noise case and (b) high noise case}
	
	\label{fig:error-histos}
\end{figure}

Going beyond a single image, we studied this system further to acquire results that could characterize the average performance of the MAP estimate for varied cases of this system. For each of the twelve possible characters known for edge dislocations in FCC aluminum, we computed the MAP estimate for $5000$ simulated experimental images. For each trial, the true dislocation position was randomly selected from within the region supported by the prior. We chose to randomize over the true dislocation position in our numerical study to capture effects that could arise based on the variation of the core's position inside a particular pixel, i.e., whether core lies near the center or the edge of a pixel. We considered a \emph{low noise} scenario where we take the noise parameters $\bar{\beta}$ and $\sigma^2$ to be $5\%$ and $1\%$ of the maximum noise-free intensity, and a \emph{high noise} scenario, where we take $\bar{\beta}$ and $\sigma^2$ to be $10\%$ and $5\%$ of the maximum noise-free intensity to demonstrate the idealized and ``worst-case'' scenarios for experimental data. \Cref{fig:error-histos} shows histograms of the estimation error of the MAP point for both noise settings. In the low noise case, we see median and mean errors of $5.3$ nm and $15.0$ nm, respectively. In the high noise case, we obtain median and mean errors of $14.2$ nm and $37.6$ nm, respectively. 

\section{Discussion}\label{sec:discussion}
\subsection{Comparison and use cases}
As demonstrated by our numerical examples, each of the three methods in this work has advantages and disadvantages that connect to different use cases for DFXM image analysis. We divide the use cases into two predominant ones described in the literature thus far. In the first, we assume a user seeks to estimate the position of a large set of dislocations from DFXM images, either for the same dislocation in a series of temporally-resolved images, or for many regions-of-interest of different dislocations across the field of view in a single DFXM image/dataset. In the second case, we assume a small number of dislocations must be characterized with the highest possible fidelity. We term these use cases the ``high-volume dislocation estimation'' case and the ``high-precision uncertainty quantification'' case, respectively.

The high-volume use case is well-suited to using the MAP point and Laplace approximation to characterize our knowledge of the position $\pos$. As shown in \cref{sec:results}, the MAP point consistently provides accurate estimates of the dislocation position in each of the test cases we consider, and takes the least computational effort of our algorithms. Also relevant to the high-volume use case, the Laplace approximation provides a relatively inexpensive method to quantify our uncertainty in the position. The computational cost of forming the Laplace approximation is simply the cost of finding the MAP point and then computing (and inverting) the Hessian at that point, which is trivial in the present problem. Hence, the computational cost is essentially the same as that of optimization. We have found that the Laplace approximation does seem to accurately capture the direction of greatest variance, measured by its alignment with computed MCMC samples. However, we have noted a phenomenon common to the Laplace approximation in our examples: the Laplace approximation seems to underestimate posterior uncertainty (see \Cref{fig:uq_results}).

By contrast, high-precision uncertainty quantification requires a more principled description of the posterior. While MCMC sampling provides the most principled description of the posterior distribution, it is also the most computationally expensive. In practice, computing posterior summaries for large data sets using MCMC sampling may necessitate high performance computing architectures. MCMC methods also require the user to compute and interpret convergence and chain mixing diagnostics like autocorrelation plots or the effective sample size of the MCMC samples \cite{brooks2011handbook}. In our numerical experiments, we saw acceptable mixing and decorrelation with a simple adaptive MCMC method, suggesting this might not be an issue for similar systems. We also note that the posterior mean is often very close to the posterior mode, and so there was little benefit in using MCMC sampling over the MAP point for position estimation. The key advantage of MCMC sampling is it fidelity to the posterior. MCMC methods provide asymptotically unbiased estimates of any posterior summary statistic and provide information on non-Gaussian features of the posterior. For example, if the posterior had multiple modes of similar strength, MCMC sampling could reveal this structure and allow the user to develop a richer, more nuanced, sense of uncertainty in the dislocation position.


\subsection{DFXM and dislocation position estimation in the context of materials science}

Understanding the motion and interactions of dislocations at the single-dislocation level in metals under various conditions is an age-old problem whose solution helps understand the microscopic origins of changes to macroscopic properties like strength, thermal transport, and phase transitions. For example, in \cite{dresselhaus2021situ}, the authors use DFXM to collect movies near the melting temperature ($T_m$) to resolve the breakdown in stability of a classical edge-dislocation (tilt) boundary in aluminum, as the thermal motion increases and interaction forces break down. That work drew its conclusions by tracking the variance in the positions of all dislocations in the boundary with slowly increasing temperatures---made possible by previous methods for dislocation tracking. In particular, the authors of \cite{dresselhaus2021situ} estimated the position of each dislocation in a temporal sequence of images using computer vision methods developed in \cite{gonzalez2020methods}. The precision of the measurements, however, was insufficient to estimate important fundamental constants like the dislocation mobility and residual stress fields. In contrast, the  algorithms developed in the present work achieve super-resolution accuracy for dislocation positions---going far beyond the $150$ nm optical resolution of the imaging system. Use of the present method offers a key opportunity to strengthen  \emph{quantitative} conclusions for these and other DFXM experiments.

We also point out the importance of our uncertainty quantification algorithms to image analysis. Using either the Laplace approximation or MCMC sampling, we obtain the directions of greatest posterior variance \emph{in the dislocation coordinate system}. Previous and ongoing work in this field has demonstrated a key attribute of high-temperature dislocation dynamics: subtle positional variance observed by DFXM indicates a combination of stochastic and locally-driven motion in the dislocations at length-scales beyond our resolution with DFXM. This has been demonstrated by \cite{dresselhaus2021situ} as essential to understand the stochastic motion of dislocations in metals at the edge of the melting temperature ($T_m$), and more recently by \cite{Guruthanan2022} in evaluating the stress fields and sub-resolution point-defect clusters necessary to describe the non-classical dislocation dynamics near $T_m$.
At temperatures near $T_m$, the density and mobility of point defects increases exponentially as the stability of the crystal breaks down. This occurs heterogeneously, as there are fluctuations in both the defect migration rates and the local temperatures in the material (i.e., hotspots). As described in \cite{cahn2001melting, dresselhaus2021situ}, these instabilities ultimately mean that each dislocation in a boundary experiences slightly different local forces based on fluctuations in the point defect density. These forces cause random-walk motion that can be detected by DFXM, which encodes information about the sub-resolution heterogeneity in the sample.
The stochastic information about the motion of these dislocations can inform the local gradients in crystalline structure, informing the local stresses being released by these high temperatures near $T_m$. As such, quantifying the uncertainty in each dislocation position \emph{in its climb and glide directions} for a range of times is essential to build the necessary models for high-$T$ dislocation mobilities that are stress invariant---an important longstanding challenge for experiments thus far.

Our position estimation and uncertainty quantification algorithms also present an important step forward for dislocation characterization in the context of the newly available subsurface dislocation tracking experiments. Over the past decade, materials science experiments have built momentum in using data-science methods to analyze microscopy results \cite{Kalinin2015}. This has enabled key opportunities for experiments to quantify information about grain-boundaries \cite{Zhou2022}, dislocations \cite{Li2020}, and other defects in statistically significant populations. Similar approaches in subsurface X-ray microscopy have struggled due to limitations in the optical resolution, and much more difficult interpretation due to the very high strain-resolution, which can make it difficult to interpret overlapping features. Measuring the positions of angstrom-sized dislocation cores with only micrometer accuracy has prevented DFXM and X-ray topography's application to many mechanical systems---in which the important length-scale for dislocation motion and dynamics is often at most a few hundred nanometers. The superresolution accuracy of our Bayesian approach demonstrates a key opportunity: to use the physics of DFXM image formation to constrain the accuracy of position estimation, allowing positional accuracy at similar scales to TEM, but over a significantly wider field of view and deep below the surface. This presents the opportunity to characterize key dislocation interactions, motion, and dynamics in sensitive regimes over which the nm-scale accuracy is essential for evaluation.


\subsection{Limitations and future work}
Our methods present the first implementation of Bayesian inference approaches in the context of DFXM; however, we note that it has limitations and avenues for improvement in its accuracy, its relevance to the high-volume use case, and refinement in addressing the DFXM and dislocation systems.

First, we comment on the high-volume use case. In this work, we manually select the prior for our synthetic evaluation images, however, a less supervised approach is clearly preferable for the high-volume use case. As noted in \cref{sec:bayes:prior}, picking the prior distribution is a modeling choice made by the user. In our numerical study, we chose our prior distribution for each image based on the shape of the dislocation in the image and our knowledge of the dislocation character and forward model. This lead to a uniform prior over a region assumed to contain the true position. Other sources of information may be incorporated into the prior distribution as well. Take for example the application considered above and presented in \cite{gonzalez2020methods, dresselhaus2021situ}, tracking the movement of dislocations through a time-series of images. In this scenario, it would be natural to use the dislocation position and velocity estimated from previous images to define the prior, rather than defining it manually. One could also use the feature extraction methods developed by \cite{gonzalez2020methods} to dictate the prior. In \cite{gonzalez2020methods}, the authors identify regions of the images associated with a dislocation using a wavelet transformation. One could take the prior to be uniform over this identified region (see \cite[Figure~3]{gonzalez2020methods} for an example). Both of the these situations lend themselves well to an automated process for defining the prior, which would be necessary to apply our methods to large data-sets of images.

A natural next step from this work would be to broaden the optimization scope described in this model. While in this work we focused on estimating the position of dislocations, the framework of Bayesian inversion may have further uses in DFXM image processing. For example, we assumed that the angles of the goniometer and the mean of the background and electric noise processes were known. An extension of our work could include these parameters in the inference problem, by defining the forward model as a function of any relevant variables (e.g., dislocation position, goniometer angles, noise parameters) and then inferring each of these using estimation algorithms analogous to algorithms presented in this work. Future progress in this work may express $\pos$ as a higher dimensional parameter to express uncertainty in all of these values. This is a natural next research direction building on the results presented in this work.

Beyond increasing the dimension of the inference problem, further progress could also be made in extending this methodology to more accurately model the optics in the microscope. The presented methodology relies heavily on the image simulation model $\forward$ and the description of measurement noise matching the true physical imaging process accurately. The success of our position estimation algorithms applied to real DFXM images will depend on the fidelity of the simulation model and noise models. More recent implementations of the forward model have been developed for X-ray free electron laser experiments \cite{Holstad2022}, and using wavefront propagation methods, which offer more accurate results (including dynamical diffraction), though at a higher computational cost \cite{Carlsen2022}. As models of the imaging process are refined, we expect the utility of the numerical methods leveraging those models to increase.

Finally, we note that the micromechanical model used within the forward model could also include nonlinear elasticity in the models. This work has used the simplest possible description of dislocations, assuming infinite dislocation lines and classical elastic theory \cite{hull2011introduction}. More complex models of dislocations have been developed---both for small well-defined dislocation boundaries \cite{Acharya2006}, and for more complex dislocation tangles or avalanches \cite{Papanikolaou2018}. Progress in adapting the DFXM forward model to more complex dislocation systems is ongoing at this time.

\section{Conclusion}\label{sec:conclusions}
This work has developed several inference methods to estimate the position of dislocations from DFXM images with quantitative uncertainty---achieving superresolution accuracy. We have done this by formulating the position estimation task as a Bayesian inference problem. We demonstrate the accuracy of these methods using different approaches and magnitudes of noise, showing that the methods obtain super-resolution accuracy for edge dislocations in FCC aluminum, with tractable uncertainty quantification in the position expressed in the dislocation system. Further advances in modeling dislocations and DFXM can be integrated into this framework directly, and our results demonstrate the robust performance of our approach. This work presents a key step forward for dislocation mechanics, as it offers positional accuracy previously only accessible via TEM. As such, our methods enable accurate multiscale characterization of \emph{subsurface} dislocations, and should allow for studies of the dynamics and interactions of dislocations across many materials systems. Future work will extend the methods presented here to higher dimensions, to infer other important parameters of interest.

\backmatter

\bmhead{Acknowledgments}
We thank Henning Friis Poulsen and Grethe Winther for their guidance and discussions as we constructed our code for the DFXM forward model used in this work.

This manuscript has been authored in part by Mission Support and Test Services, LLC, under Contract No. DE-NA0003624 with the U.S. Department of Energy and supported by the Site-Directed Research and Development Program, U.S. Department of Energy, National Nuclear Security Administration. The United States Government retains and the publisher, by accepting the article for publication, acknowledges that the United States Government retains a non-exclusive, paid-up, irrevocable, worldwide license to publish or reproduce the published form of this manuscript, or allow others to do so, for United States Government purposes. The U.S. Department of Energy will provide public access to these results of federally sponsored research in accordance with the DOE Public Access Plan (http://energy.gov/downloads/doe-public-access-plan). The views expressed in the article do not necessarily represent the views of the U.S. Department of Energy or the United States Government. DOE/NV/03624--1304

LEDM's intial contributions were performed under the auspices of the U.S. Department of Energy by Lawrence Livermore National Laboratory under Contract DE-AC52-07NA27344, and the Lawrence Fellowship. All analysis and writing were performed at Stanford University and SLAC National Accelerator Labs.


\section*{Declarations}

The authors declare no conflicts of interest.


\bibliography{dfxm}


\begin{thebibliography}{31}
\ifx \bisbn   \undefined \def \bisbn  #1{ISBN #1}\fi
\ifx \binits  \undefined \def \binits#1{#1}\fi
\ifx \bauthor  \undefined \def \bauthor#1{#1}\fi
\ifx \batitle  \undefined \def \batitle#1{#1}\fi
\ifx \bjtitle  \undefined \def \bjtitle#1{#1}\fi
\ifx \bvolume  \undefined \def \bvolume#1{\textbf{#1}}\fi
\ifx \byear  \undefined \def \byear#1{#1}\fi
\ifx \bissue  \undefined \def \bissue#1{#1}\fi
\ifx \bfpage  \undefined \def \bfpage#1{#1}\fi
\ifx \blpage  \undefined \def \blpage #1{#1}\fi
\ifx \burl  \undefined \def \burl#1{\textsf{#1}}\fi
\ifx \doiurl  \undefined \def \doiurl#1{\url{https://doi.org/#1}}\fi
\ifx \betal  \undefined \def \betal{\textit{et al.}}\fi
\ifx \binstitute  \undefined \def \binstitute#1{#1}\fi
\ifx \binstitutionaled  \undefined \def \binstitutionaled#1{#1}\fi
\ifx \bctitle  \undefined \def \bctitle#1{#1}\fi
\ifx \beditor  \undefined \def \beditor#1{#1}\fi
\ifx \bpublisher  \undefined \def \bpublisher#1{#1}\fi
\ifx \bbtitle  \undefined \def \bbtitle#1{#1}\fi
\ifx \bedition  \undefined \def \bedition#1{#1}\fi
\ifx \bseriesno  \undefined \def \bseriesno#1{#1}\fi
\ifx \blocation  \undefined \def \blocation#1{#1}\fi
\ifx \bsertitle  \undefined \def \bsertitle#1{#1}\fi
\ifx \bsnm \undefined \def \bsnm#1{#1}\fi
\ifx \bsuffix \undefined \def \bsuffix#1{#1}\fi
\ifx \bparticle \undefined \def \bparticle#1{#1}\fi
\ifx \barticle \undefined \def \barticle#1{#1}\fi
\bibcommenthead
\ifx \bconfdate \undefined \def \bconfdate #1{#1}\fi
\ifx \botherref \undefined \def \botherref #1{#1}\fi
\ifx \url \undefined \def \url#1{\textsf{#1}}\fi
\ifx \bchapter \undefined \def \bchapter#1{#1}\fi
\ifx \bbook \undefined \def \bbook#1{#1}\fi
\ifx \bcomment \undefined \def \bcomment#1{#1}\fi
\ifx \oauthor \undefined \def \oauthor#1{#1}\fi
\ifx \citeauthoryear \undefined \def \citeauthoryear#1{#1}\fi
\ifx \endbibitem  \undefined \def \endbibitem {}\fi
\ifx \bconflocation  \undefined \def \bconflocation#1{#1}\fi
\ifx \arxivurl  \undefined \def \arxivurl#1{\textsf{#1}}\fi
\csname PreBibitemsHook\endcsname

\bibitem{hull2011introduction}
\begin{bbook}
\bauthor{\bsnm{Hull}, \binits{D.}},
\bauthor{\bsnm{Bacon}, \binits{D.J.}}:
\bbtitle{Introduction to Dislocations}
vol. \bseriesno{37}.
\bpublisher{Elsevier},
\blocation{The Boulevard, Langford Lane, Kidlington, Oxford, OX5 1GB}
(\byear{2011})
\end{bbook}
\endbibitem

\bibitem{Kubin2011}
\begin{barticle}
\bauthor{\bsnm{Sauzay}, \binits{M.}},
\bauthor{\bsnm{Kubin}, \binits{L.P.}}:
\batitle{Scaling laws for dislocation microstructures in monotonic and cyclic
  deformation of {FCC} metals}.
\bjtitle{Progress in Materials Science}
\bvolume{56},
\bfpage{725}--\blpage{784}
(\byear{2011})
\end{barticle}
\endbibitem

\bibitem{Danilewsky2020}
\begin{barticle}
\bauthor{\bsnm{Danilewsky}, \binits{A.N.}}:
\batitle{{X}-ray topography—more than nice pictures}.
\bjtitle{Crystal Research and Technology}
\bvolume{55}(\bissue{9}),
\bfpage{2000012}
(\byear{2020})
\end{barticle}
\endbibitem

\bibitem{kutsal2019esrf}
\begin{bchapter}
\bauthor{\bsnm{Kutsal}, \binits{M.}},
\bauthor{\bsnm{Bernard}, \binits{P.}},
\bauthor{\bsnm{Berruyer}, \binits{G.}},
\bauthor{\bsnm{Cook}, \binits{P.}},
\bauthor{\bsnm{Hino}, \binits{R.}},
\bauthor{\bsnm{Jakobsen}, \binits{A.}},
\bauthor{\bsnm{Ludwig}, \binits{W.}},
\bauthor{\bsnm{Ormstrup}, \binits{J.}},
\bauthor{\bsnm{Roth}, \binits{T.}},
\bauthor{\bsnm{Simons}, \binits{H.}}, \betal:
\bctitle{The {ESRF} dark-field {X}-ray microscope at {ID}06}.
In: \bbtitle{IOP Conference Series: Materials Science and Engineering},
vol. \bseriesno{580},
p. \bfpage{012007}
(\byear{2019}).
\bcomment{IOP Publishing}
\end{bchapter}
\endbibitem

\bibitem{simons2015dark}
\begin{barticle}
\bauthor{\bsnm{Simons}, \binits{H.}},
\bauthor{\bsnm{King}, \binits{A.}},
\bauthor{\bsnm{Ludwig}, \binits{W.}},
\bauthor{\bsnm{Detlefs}, \binits{C.}},
\bauthor{\bsnm{Pantleon}, \binits{W.}},
\bauthor{\bsnm{Schmidt}, \binits{S.}},
\bauthor{\bsnm{St{\"o}hr}, \binits{F.}},
\bauthor{\bsnm{Snigireva}, \binits{I.}},
\bauthor{\bsnm{Snigirev}, \binits{A.}},
\bauthor{\bsnm{Poulsen}, \binits{H.F.}}:
\batitle{Dark-field {X}-ray microscopy for multiscale structural
  characterization}.
\bjtitle{Nature communications}
\bvolume{6}(\bissue{1}),
\bfpage{1}--\blpage{6}
(\byear{2015})
\end{barticle}
\endbibitem

\bibitem{poulsen2017x}
\begin{barticle}
\bauthor{\bsnm{Poulsen}, \binits{H.F.}},
\bauthor{\bsnm{Jakobsen}, \binits{A.}},
\bauthor{\bsnm{Simons}, \binits{H.}},
\bauthor{\bsnm{Ahl}, \binits{S.R.}},
\bauthor{\bsnm{Cook}, \binits{P.}},
\bauthor{\bsnm{Detlefs}, \binits{C.}}:
\batitle{{X}-ray diffraction microscopy based on refractive optics}.
\bjtitle{Journal of Applied Crystallography}
\bvolume{50}(\bissue{5}),
\bfpage{1441}--\blpage{1456}
(\byear{2017})
\end{barticle}
\endbibitem

\bibitem{poulsen2020multi}
\begin{barticle}
\bauthor{\bsnm{Poulsen}, \binits{H.F.}}:
\batitle{Multiscale hard {X}-ray microscopy}.
\bjtitle{Current Opinion in Solid State and Materials Science}
\bvolume{24}(\bissue{2}),
\bfpage{100820}
(\byear{2020})
\end{barticle}
\endbibitem

\bibitem{simons2018long}
\begin{barticle}
\bauthor{\bsnm{Simons}, \binits{H.}},
\bauthor{\bsnm{Haugen}, \binits{A.B.}},
\bauthor{\bsnm{Jakobsen}, \binits{A.C.}},
\bauthor{\bsnm{Schmidt}, \binits{S.}},
\bauthor{\bsnm{St{\"o}hr}, \binits{F.}},
\bauthor{\bsnm{Majkut}, \binits{M.}},
\bauthor{\bsnm{Detlefs}, \binits{C.}},
\bauthor{\bsnm{Daniels}, \binits{J.E.}},
\bauthor{\bsnm{Damjanovic}, \binits{D.}},
\bauthor{\bsnm{Poulsen}, \binits{H.F.}}:
\batitle{Long-range symmetry breaking in embedded ferroelectrics}.
\bjtitle{Nature materials}
\bvolume{17}(\bissue{9}),
\bfpage{814}--\blpage{819}
(\byear{2018})
\end{barticle}
\endbibitem

\bibitem{cook2018insights}
\begin{barticle}
\bauthor{\bsnm{Cook}, \binits{P.K.}},
\bauthor{\bsnm{Simons}, \binits{H.}},
\bauthor{\bsnm{Jakobsen}, \binits{A.C.}},
\bauthor{\bsnm{Yildirim}, \binits{C.}},
\bauthor{\bsnm{Poulsen}, \binits{H.F.}},
\bauthor{\bsnm{Detlefs}, \binits{C.}}:
\batitle{Insights into the exceptional crystallographic order of biominerals
  using dark-field {X}-ray microscopy}.
\bjtitle{Microscopy and Microanalysis}
\bvolume{24}(\bissue{S2}),
\bfpage{88}--\blpage{89}
(\byear{2018})
\end{barticle}
\endbibitem

\bibitem{jakobsen2019mapping}
\begin{barticle}
\bauthor{\bsnm{Jakobsen}, \binits{A.}},
\bauthor{\bsnm{Simons}, \binits{H.}},
\bauthor{\bsnm{Ludwig}, \binits{W.}},
\bauthor{\bsnm{Yildirim}, \binits{C.}},
\bauthor{\bsnm{Leemreize}, \binits{H.}},
\bauthor{\bsnm{Porz}, \binits{L.}},
\bauthor{\bsnm{Detlefs}, \binits{C.}},
\bauthor{\bsnm{Poulsen}, \binits{H.}}:
\batitle{Mapping of individual dislocations with dark-field {X}-ray
  microscopy}.
\bjtitle{Journal of Applied Crystallography}
\bvolume{52}(\bissue{1}),
\bfpage{122}--\blpage{132}
(\byear{2019})
\end{barticle}
\endbibitem

\bibitem{poulsen2020geometrical}
\begin{botherref}
\oauthor{\bsnm{Poulsen}, \binits{H.}},
\oauthor{\bsnm{Dresselhaus-Marais}, \binits{L.}},
\oauthor{\bsnm{Carlsen}, \binits{M.}},
\oauthor{\bsnm{Detlefs}, \binits{C.}},
\oauthor{\bsnm{Winther}, \binits{G.}}:
Geometrical-optics formalism to model contrast in dark-field {X}-ray
  microscopy.
Journal of Applied Crystallography
\textbf{54}(6)
(2021)
\end{botherref}
\endbibitem

\bibitem{gonzalez2020methods}
\begin{botherref}
\oauthor{\bsnm{Gonzalez}, \binits{A.}},
\oauthor{\bsnm{Howard}, \binits{M.}},
\oauthor{\bsnm{Breckling}, \binits{S.}},
\oauthor{\bsnm{Dresselhaus-Marais}, \binits{L.E.}}:
Methods to quantify dislocation behavior with dark-field {X}-ray microscopy
  timescans of single-crystal aluminum.
arXiv preprint arXiv:2008.04972
(2020)
\end{botherref}
\endbibitem

\bibitem{dresselhaus2021situ}
\begin{barticle}
\bauthor{\bsnm{Dresselhaus-Marais}, \binits{L.E.}},
\bauthor{\bsnm{Winther}, \binits{G.}},
\bauthor{\bsnm{Howard}, \binits{M.}},
\bauthor{\bsnm{Gonzalez}, \binits{A.}},
\bauthor{\bsnm{Breckling}, \binits{S.R.}},
\bauthor{\bsnm{Yildirim}, \binits{C.}},
\bauthor{\bsnm{Cook}, \binits{P.K.}},
\bauthor{\bsnm{Kutsal}, \binits{M.}},
\bauthor{\bsnm{Simons}, \binits{H.}},
\bauthor{\bsnm{Detlefs}, \binits{C.}}, \betal:
\batitle{In situ visualization of long-range defect interactions at the edge of
  melting}.
\bjtitle{Science Advances}
\bvolume{7}(\bissue{29}),
\bfpage{8311}
(\byear{2021})
\end{barticle}
\endbibitem

\bibitem{rossi2006pixel}
\begin{bbook}
\bauthor{\bsnm{Rossi}, \binits{L.}},
\bauthor{\bsnm{Fischer}, \binits{P.}},
\bauthor{\bsnm{Rohe}, \binits{T.}},
\bauthor{\bsnm{Wermes}, \binits{N.}}:
\bbtitle{Pixel Detectors: From Fundamentals to Applications}.
\bpublisher{Springer},
\blocation{Verlag Berlin Heidelberg}
(\byear{2006})
\end{bbook}
\endbibitem

\bibitem{snyder1993image}
\begin{barticle}
\bauthor{\bsnm{Snyder}, \binits{D.L.}},
\bauthor{\bsnm{Hammoud}, \binits{A.M.}},
\bauthor{\bsnm{White}, \binits{R.L.}}:
\batitle{Image recovery from data acquired with a charge-coupled-device
  camera}.
\bjtitle{JOSA A}
\bvolume{10}(\bissue{5}),
\bfpage{1014}--\blpage{1023}
(\byear{1993})
\end{barticle}
\endbibitem

\bibitem{snyder1995compensation}
\begin{barticle}
\bauthor{\bsnm{Snyder}, \binits{D.L.}},
\bauthor{\bsnm{Helstrom}, \binits{C.W.}},
\bauthor{\bsnm{Lanterman}, \binits{A.D.}},
\bauthor{\bsnm{Faisal}, \binits{M.}},
\bauthor{\bsnm{White}, \binits{R.L.}}:
\batitle{Compensation for readout noise in {CCD} images}.
\bjtitle{JOSA A}
\bvolume{12}(\bissue{2}),
\bfpage{272}--\blpage{283}
(\byear{1995})
\end{barticle}
\endbibitem

\bibitem{mogensen2018optim}
\begin{barticle}
\bauthor{\bsnm{Mogensen}, \binits{P.K.}},
\bauthor{\bsnm{Riseth}, \binits{A.N.}}:
\batitle{Optim: A mathematical optimization package for {Julia}}.
\bjtitle{Journal of Open Source Software}
\bvolume{3}(\bissue{24}),
\bfpage{615}
(\byear{2018}).
\doiurl{10.21105/joss.00615}
\end{barticle}
\endbibitem

\bibitem{raftery1996approximate}
\begin{barticle}
\bauthor{\bsnm{Raftery}, \binits{A.E.}}:
\batitle{Approximate {B}ayes factors and accounting for model uncertainty in
  generalised linear models}.
\bjtitle{Biometrika}
\bvolume{83}(\bissue{2}),
\bfpage{251}--\blpage{266}
(\byear{1996})
\end{barticle}
\endbibitem

\bibitem{brooks2011handbook}
\begin{bbook}
\bauthor{\bsnm{Brooks}, \binits{S.}},
\bauthor{\bsnm{Gelman}, \binits{A.}},
\bauthor{\bsnm{Jones}, \binits{G.}},
\bauthor{\bsnm{Meng}, \binits{X.-L.}}:
\bbtitle{Handbook of Markov {chain} Monte Carlo}.
\bpublisher{CRC press},
\blocation{6000 Broken Sound Parkway NW, Suite 300, Boca Raton, FL 33487-2742}
(\byear{2011})
\end{bbook}
\endbibitem

\bibitem{haario2001adaptive}
\begin{barticle}
\bauthor{\bsnm{Haario}, \binits{H.}},
\bauthor{\bsnm{Saksman}, \binits{E.}},
\bauthor{\bsnm{Tamminen}, \binits{J.}}, \betal:
\batitle{An adaptive {Metropolis} algorithm}.
\bjtitle{Bernoulli}
\bvolume{7}(\bissue{2}),
\bfpage{223}--\blpage{242}
(\byear{2001})
\end{barticle}
\endbibitem

\bibitem{vihola2014ergonomic}
\begin{botherref}
\oauthor{\bsnm{Vihola}, \binits{M.}}:
Ergonomic and reliable {B}ayesian inference with adaptive {Markov chain Monte
  Carlo}.
Wiley stats Ref: statistics reference online,
1--12
(2014)
\end{botherref}
\endbibitem

\bibitem{Yildirim2022}
\begin{botherref}
\oauthor{\bsnm{Yildirim}, \binits{C.}},
\oauthor{\bsnm{Winther}, \binits{G.}},
\oauthor{\bsnm{Poulsen}, \binits{H.F.}},
\oauthor{\bsnm{Detlefs}, \binits{C.}},
\oauthor{\bsnm{Huang}, \binits{P.-H.}},
\oauthor{\bsnm{Dresselhaus-Marais}, \binits{L.E.}}:
Extensive 3{D} mapping of dislocation boundaries by dark-field {X}-ray
  microscopy.
In Preparation
(2022)
\end{botherref}
\endbibitem

\bibitem{Guruthanan2022}
\begin{botherref}
\oauthor{\bsnm{Guruthanan}, \binits{R.}},
\oauthor{\bsnm{Dresselhaus-Marais}, \binits{L.E.}}:
Vibrating string model to describe boundary migration near {$T_m$}.
In Preparation
(2022)
\end{botherref}
\endbibitem

\bibitem{cahn2001melting}
\begin{barticle}
\bauthor{\bsnm{Cahn}, \binits{R.W.}}:
\batitle{Melting from within}.
\bjtitle{Nature}
\bvolume{413}(\bissue{6856}),
\bfpage{582}--\blpage{583}
(\byear{2001})
\end{barticle}
\endbibitem

\bibitem{Kalinin2015}
\begin{botherref}
\oauthor{\bsnm{Kalinin}, \binits{S.V.}},
\oauthor{\bsnm{Sumpter}, \binits{B.G.}},
\oauthor{\bsnm{Archibald}, \binits{R.K.}}:
Big–deep–smart data in imaging for guiding materials design.
Nature Materials,
973--980
(2015)
\end{botherref}
\endbibitem

\bibitem{Zhou2022}
\begin{barticle}
\bauthor{\bsnm{Zhou}, \binits{X.}},
\bauthor{\bsnm{Wei}, \binits{Y.}},
\bauthor{\bsnm{Kühbach}, \binits{M.}},
\bauthor{\bsnm{Zhao}, \binits{H.}},
\bauthor{\bsnm{Vogeld}, \binits{F.}},
\bauthor{\bsnm{Kamachali}, \binits{R.D.}},
\bauthor{\bsnm{Thompson}, \binits{G.B.}},
\bauthor{\bsnm{Raabe}, \binits{D.}},
\bauthor{\bsnm{Gaulta}, \binits{B.}}:
\batitle{Revealing in-plane grain boundary composition features through machine
  learning from atom probe tomography data}.
\bjtitle{Acta Materialia}
\bvolume{220},
\bfpage{117633}
(\byear{2022})
\end{barticle}
\endbibitem

\bibitem{Li2020}
\begin{barticle}
\bauthor{\bsnm{Li}, \binits{W.}},
\bauthor{\bsnm{Field}, \binits{K.G.}},
\bauthor{\bsnm{Morgan}, \binits{D.}}:
\batitle{Automated defect analysis in electron microscopic images}.
\bjtitle{npj Comput Mater}
\bvolume{4},
\bfpage{36}
(\byear{2018})
\end{barticle}
\endbibitem

\bibitem{Holstad2022}
\begin{botherref}
\oauthor{\bsnm{Holstad}, \binits{T.S.}},
\oauthor{\bsnm{R{\ae}der}, \binits{T.M.}},
\oauthor{\bsnm{Carlsen}, \binits{M.}},
\oauthor{\bsnm{Bergb{\"a}ck~Knudsen}, \binits{E.}},
\oauthor{\bsnm{Dresselhaus-Marais}, \binits{L.E.}},
\oauthor{\bsnm{Haldrup}, \binits{K.}},
\oauthor{\bsnm{Simons}, \binits{H.}},
\oauthor{\bsnm{Nielsen}, \binits{M.M.}},
\oauthor{\bsnm{Poulsen}, \binits{H.F.}}:
{X}-ray free-electron laser based dark-field {X}-ray microscopy: a
  simulation-based study.
Journal of Applied Crystallography
\textbf{55}(1)
(2022)
\end{botherref}
\endbibitem

\bibitem{Carlsen2022}
\begin{botherref}
\oauthor{\bsnm{Carlsen}, \binits{M.}},
\oauthor{\bsnm{Detlefs}, \binits{C.}},
\oauthor{\bsnm{Yildirim}, \binits{C.}},
\oauthor{\bsnm{Ræder}, \binits{T.}},
\oauthor{\bsnm{Simons}, \binits{H.}}:
Simulating dark-field {X}-ray microscopy images with wave front propagation
  techniques.
arXiv:2201.07549
(2022)
\end{botherref}
\endbibitem

\bibitem{Acharya2006}
\begin{barticle}
\bauthor{\bsnm{Acharya}, \binits{A.}},
\bauthor{\bsnm{Roy}, \binits{A.}}:
\batitle{Size effects and idealized dislocation microstructure at small scales:
  Predictions of a phenomenological model of mesoscopic field dislocation
  mechanics: Part i}.
\bjtitle{Journal of the Mechanics and Physics of Solids}
\bvolume{54},
\bfpage{1687}--\blpage{1710}
(\byear{2006})
\end{barticle}
\endbibitem

\bibitem{Papanikolaou2018}
\begin{barticle}
\bauthor{\bsnm{Papanikolaou}, \binits{S.}},
\bauthor{\bsnm{Cui}, \binits{Y.}},
\bauthor{\bsnm{Ghoniem}, \binits{N.}}:
\batitle{Avalanches and plastic flow in crystal plasticity: an overview}.
\bjtitle{Modelling and Simulation in Materials Science and Engineering}
\bvolume{26},
\bfpage{013001}
(\byear{2018})
\end{barticle}
\endbibitem

\end{thebibliography}

\end{document}